\documentclass[lettersize,journal]{IEEEtran}
\usepackage[utf8]{inputenc}  
\usepackage{amsmath}
\usepackage{array}
\usepackage[caption=false,font=normalsize,labelfont=sf,textfont=sf]{subfig}
\usepackage{textcomp}
\usepackage{stfloats}
\usepackage{url}
\usepackage{verbatim}
\usepackage{graphicx}
\usepackage{cite}
\usepackage{booktabs}
\usepackage{tabulary}
\usepackage{tabularx}
\usepackage{longtable}
\usepackage{multirow}
\usepackage{float}
\usepackage{hyperref}
\hypersetup{
    colorlinks=true,
    linkcolor=blue,
    filecolor=magenta,      
    urlcolor=cyan,
    pdftitle={Overleaf Example},
    pdfpagemode=FullScreen,}

\begin{document}

\title{Identity Card Presentation Attack Detection: A Systematic Review}

\author{
    Esteban M. Ruiz, Juan~E.~Tapia,~\IEEEmembership{Senior Member,~IEEE,} Reinel T.~Soto, and~Christoph~Busch,~\IEEEmembership{Senior~Member,~IEEE}%
\thanks{Juan Tapia and Christoph Busch are with the da/sec-Biometrics and Internet Security Research Group, Hochschule Darmstadt, Germany, e-mail: \{\href{juan.tapia-farias@h-da.de}{juan.tapia-farias}, \href{christoph.busch@h-da.de}{christoph.busch}\}@h-da.de.}%

\thanks{Esteban M. Ruiz is with the Universidad Autónoma de Manizales, email: \href{mailto:esteban.mercador@autonoma.edu.co}{esteban.mercador@autonoma.edu.co}.}
\thanks{Reinel T. Soto is with the Universidad Autónoma de Manizales and Universidad de Caldas, email: \href{mailto:rtabares@autonoma.edu.co}{rtabares@autonoma.edu.co}.}
\thanks{Manuscript received Month DD, YYYY; revised Month DD, YYYY.}}


\maketitle
\begin{abstract}
Remote identity verification is essential for modern digital security; however, it remains highly vulnerable to sophisticated Presentation Attacks (PAs) that utilise forged or manipulated identity documents. Although Deep Learning (DL) has driven advances in Presentation Attack Detection (PAD), the field is fundamentally limited by a lack of data and the poor generalisation of models across various document types and new attack methods.

This article presents a systematic literature review (SLR) conducted in accordance with the PRISMA methodology, aiming to analyse and synthesise the current state of AI-based PAD for identity documents from 2020 to 2025 comprehensively. Our analysis reveals a significant methodological evolution: a transition from standard Convolutional Neural Networks (CNNs) to specialised forensic micro-artefact analysis, and more recently, the adoption of large-scale Foundation Models (FMs), marking a substantial shift in the field.

We identify a central paradox that hinders progress: a critical "Reality Gap" exists between models validated on extensive, private datasets and those assessed using limited public datasets, which typically consist of mock-ups or synthetic data. This gap limits the reproducibility of research results. Additionally, we highlight a "Synthetic Utility Gap," where synthetic data generation — the primary academic response to data scarcity — often fails to predict forensic utility. This can lead to model overfitting to generation artefacts instead of the actual attack. 

This review consolidates our findings, identifies critical research gaps, and provides a definitive reference framework that outlines a prescriptive roadmap for future research aimed at developing secure, robust, and globally generalizable PAD systems.
\end{abstract}


\section{Introduction}
\begin{figure*}[]
    \centering
    \includegraphics[width=0.85\linewidth]{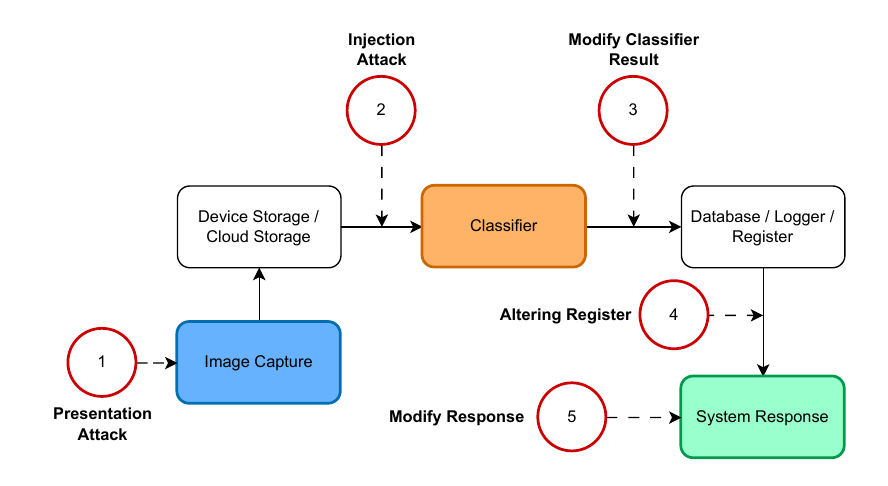}
    \caption{Probable points of attack in a typical identity document verification workflow.}
    \label{fig:attack_points}
\end{figure*}

Identity verification is a fundamental process in numerous digital security systems, playing a crucial role in authenticating individuals, particularly in the increasingly prevalent context of \textbf{remote onboarding processes}. The identity verification process, while seemingly straightforward, is a multi-stage workflow with several points of vulnerability. As illustrated in Figure \ref{fig:attack_points}, malicious actors can target the system at various stages of processing. These attacks include: \textbf{(1) Presentation Attacks (PAs)}, where a physically falsified or manipulated document is presented to the capture device; \textbf{(2) Injection Attacks}, where imperceptible noise is injected into the digital image before classification to mislead the model; \textbf{(3) Modify Classifier Result}, which involve compromising the classifier to directly alter its output scores; \textbf{(4) Altering Register}, where transaction logs are illicitly modified to change the verification outcome; and \textbf{(5) Modify Response}, targeting the final decision logic to force an illegitimate approval.

While acknowledging the critical threat posed by each of these vectors, \textbf{this survey focuses specifically on Presentation Attacks (Input 1)}. PAs represent the most common and accessible threat vector in remote onboarding scenarios, as they exploit the initial interaction between the user and the system and do not necessarily require advanced digital intrusion or penetration skills.

The Presentation Attack Detection (PAD) module takes part in this process that typically relies on users providing a self-captured image of their face (a selfie) and a \textbf{captured image of their identity document}, which are then submitted for an automated validation and biometric verification process. These processes are employed in financial, governmental, and commercial services and rely on automatic detection mechanisms to ensure identity, which serves as trust anchor. However, PAs on identity documents (ID) have evolved significantly in recent years, posing a growing threat to these systems. PAs include techniques such as printed, screen, and composite/border attacks, which represent manual or digital manipulation of an ID, all designed to deceive identity validation systems \cite{benalcazar_synthetic_2023}. 

Traditional methods of document verification have proven weak against increasingly sophisticated presentation attacks. The implementation of additional document security features, such as holographic elements, watermarks, and special inks, has made document forgery more complicated. Nevertheless, technological advances have enabled the synthetic replication of these security measures, requiring constant evolution in fraud detection systems. Given this scenario, automatic verification systems must adapt to address the emerging challenges posed by attackers, including those in remote verification systems.

Furthermore, presentation attacks not only compromise the security of authentication systems, but also have significant legal and economic implications. Financial and governmental institutions have reported substantial losses due to the proliferation of forged identity documents. In the United States alone, reported losses from fraud surpassed \$10 billion in 2023, with bank transfers and payments being the highest loss method, indicating direct institutional impact \cite{noauthor_as_2024}. 

Beyond these institutional repercussions, the failure to detect a presentation attack is often the genesis of identity fraud, a crime with devastating consequences for individual victims. A comprehensive study conducted in 2024 \cite{irvin-erickson_identity_2024} highlights that victims of identity fraud experience a wide range of harms that extend far beyond direct financial losses. These harms include significant emotional distress, damage to their personal reputation, and a substantial time burden spent resolving the aftermath of their victimisation. Therefore, the development of robust PAD systems is not merely a matter of corporate security but also a critical preventative measure to shield individuals from the severe personal damages associated with this crime.

A critical impediment to developing robust PAD solutions is the pronounced scarcity of publicly available, large-scale datasets containing bona fide ID-card samples and attack samples. This scarcity is largely attributable to the stringent General Data Protection Regulation (GDPR) and legitimate privacy concerns associated with collecting and disseminating images of both bona fide and fraudulent identity documents, especially those used in actual attacks. However, research into presentation attack detection methods is crucial for enhancing security and reliability in authentication processes and must evolve despite the scare data availability. An example of the onboarding process, which highlights the PAD on ID-cards is illustrated in Figure \ref{fig:onboarding}:

\begin{figure*}[]
    \centering
    \includegraphics[width=0.8\linewidth]{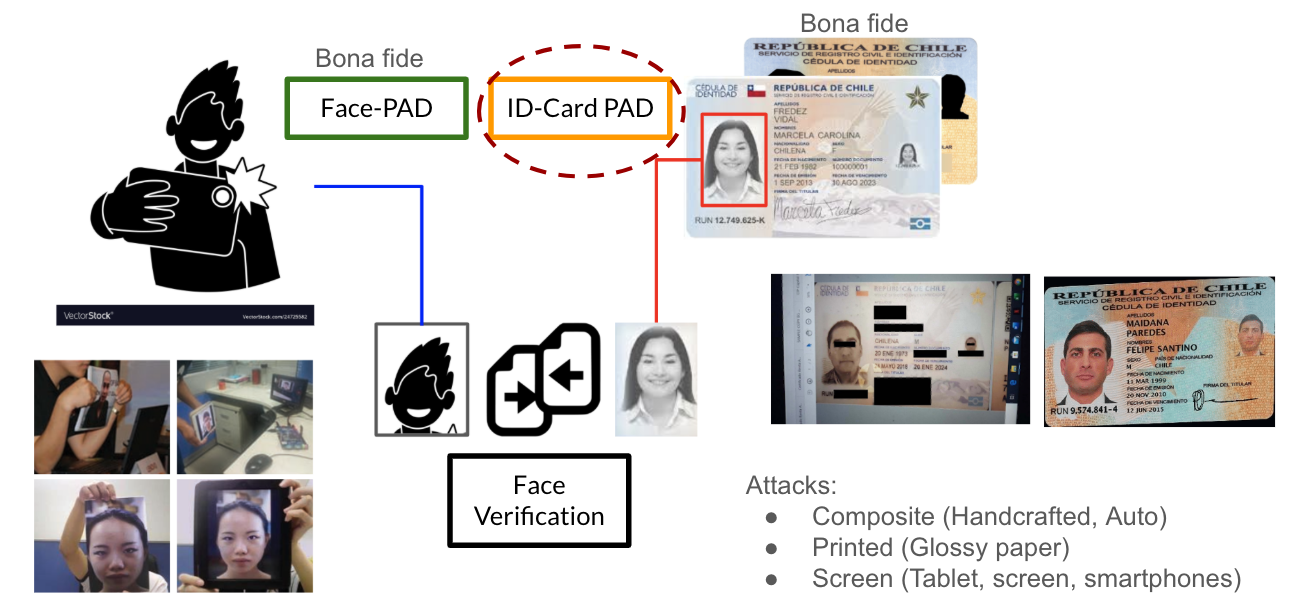}
    \caption{PAD functionality within an onboarding process.}
    \label{fig:onboarding}
\end{figure*}

Currently, in the state of the art, we have identified different kinds of bona fide samples and attack samples, that can be found in image databases used for PAD research. Based on that, we can define the most common scenarios:

\begin{itemize}
    \item \textbf{Bona fide:} Pristine image captured from a genuine ID-card officially issued by the government entity, which is presented physically in front of the capture device at the moment of the onboarding or verification process.

    \item \textbf{Printed:} Type of physical Presentation Attack (PA) where a copy of the identity document is printed on a substrate, commonly paper, which is than acquired by the capture device (e.g., smartphone). These attacks can vary in quality, ranging from prints on plain or glossy paper to high-resolution reproductions with a large number of dots per inch (dpi).

    \item \textbf{Screen:} Type of physical Presentation Attack (PA), also known as a recapture or replay attack, this consists of displaying a digital image of an identity document on an electronic screen (e.g., smartphone, tablet, monitor) and capturing it with the capture device (e.g., smartphone).

    \item \textbf{Composite/Border:} This refers to a physical manipulation attack in which the content of a physical identity document image is modified prior to the capture process. Common techniques include splicing (cutting and pasting regions, such as photographs or signatures, from another document), inpainting (erasing and rewriting text fields), or face morphing (merging two faces into a portrait photo). A digital composite version is also possible. This term encompasses attacks where the identity document is created from scratch using specialised image editing tools or AI tools. The resulting document is then printed onto PVC sheets or similar materials to simulate a real, physical document. The modified document is eventually captured by the capture device.
    
    \item \textbf{Simulated bona fide :} This type of document is created using generative AI, specifically Generative Adversarial Networks, Stable Diffusion, and other similar tools. This simulated bona fide document has been created in order to mimic a bona fide image, which has been demonstrated of equivalent distribution of derived (i.e., computed) features. On this condition, we anticipate that the PA classifier assigns samples of this type to the class of bona fide images.
\end{itemize}

The robustness and generalisation capability of any PAD system, particularly those based on Deep Learning, are intrinsically dependent on the number of images, quality of signal, diversity of presentation attack instruments, and representativeness of the datasets employed for their training and the data used in the evaluation. The composition of these datasets are fundamental not only for developing models capable of discerning between bona fide and attack presentations but also for benchmarking different methodological approaches. Nevertheless, the creation and availability of large-scale public datasets with variability in presentation attacks on identity documents remain a significant challenge, often limiting the reproducibility of studies and collaborative progress in the field for both academia and companies. 

This study reviews the databases currently available used in the state of the art and the last algorithms utilised in PAD research for identity documents (Section \ref{sec:datasets}), also highlighting efforts in synthetic data generation to mitigate issues such as class imbalance and the aforementioned scarcity of bona fide and attack samples.

The appropriate selection of evaluation metrics is also crucial for quantifying the effectiveness of models in attack detection, as well as for evaluating the similarity of synthetic images, in order to understand their limitations and potential biases. In this review, commonly employed performance metrics based on ISO/IEC 30107-3 \cite{ISO-IEC-30107-3-PAD-metrics-2023} for PAD, such as the Attack Presentation Classification Error Rate (APCER) and the Bona Fide Presentation Classification Error Rate (BPCER), are analysed throughout this work. 
Moreover, as metrics to assess the quality of synthetic images generation we use the Fréchet Inception Distance (FID)\cite{FID}, Learned Perceptual Image Patch Similarity (LPIPS) \cite{LPIPS} and VGG-Loss \cite{VGGloss}. Furthermore, the importance of adhering to standardised evaluation protocols (Section \ref{sec:metrics}) is discussed to ensure the comparability and reproducibility of results reported in the literature.

In summary, the contributions of this survey are:

\begin{itemize}
    \item This survey was conducted in accordance with the Preferred Reporting Items for Systematic Reviews and Meta-Analyses (PRISMA) methodology using a systematic literature review \cite{page_prisma_2021}.
    \item A survey of more than 29 PAD studies and datasets on ID-cards from 2020 to 2025 (Section \ref{sec:datasets} and \ref{sec:pad_methods}), including condensed overview tables for the reviewed publications and their datasets (Table \ref{tab:datasets_synthesis})
    \item A detailed discussion of various PAD approaches for ID-cards, issues and challenges (Section \ref{sec:discussion}), including avenues for future work.
\end{itemize}

\section{Systematic Literature Review}
\label{sec:slr}

To ensure a comprehensive, transparent, and reproducible analysis of the state-of-the-art in PAD for identity documents, this study employs a systematic literature review (SLR) methodology. The process is structured following the principles outlined in the PRISMA 2020 framework \cite{page_prisma_2021, Moherb2535}. The primary objective is to identify, evaluate, and synthesise recent research on methods that are based on Deep Learning (DL) and Foundation Models (FM) for both detecting presentation attacks and generating synthetic data. At the same time, we address the prevalent challenges of data scarcity, imbalance, and privacy protection.

\subsection{Research Questions}
\label{subsec:rqs}

To provide a clear structure for this review and a robust basis for the data extraction process, we defined four specific Research Questions (RQs). These RQs are designed to map the landscape of available tools, current methodologies, evaluation standards, and the key challenges that define the field:

\begin{itemize}
    \item \textbf{RQ1 (Datasets):} What publicly available datasets exist for ID-card PAD, what are their scopes (document types, bona fide nature), and what specific attack types do they model?
    \item \textbf{RQ2 (Architectures):} What are the primary Deep Learning (DL) and Foundation Model (FM) architectures and training paradigms proposed for detecting presentation attacks on identity documents?
    \item \textbf{RQ3 (Performance Metrics):} What are the standard performance metrics reported for evaluating PAD system efficacy?
    \item \textbf{RQ4 (Synthetic Data \& Gaps):} How are synthetic data generation methods being used to address data scarcity, how is their quality evaluated and what are the principal identified research gaps?
\end{itemize}

The remainder of this review is structured to explicitly answer these questions, with dedicated sections analyzing datasets (RQ1, Section \ref{sec:datasets}), methodologies (RQ2, Section \ref{sec:pad_methods}), and metrics (RQ3, Section \ref{sec:metrics}), culminating in a discussion that synthesizes the answers to RQ4 (Section \ref{sec:discussion}).

\subsection{Search Strategy}
\label{subsec:search_strategy}

A systematic search was conducted in July 2025 across five major academic databases to ensure wide coverage of the relevant literature. The primary databases selected for their extensive indexing in computer science, engineering, and biometrics were: \textbf{Scopus, ACM Digital Library, SpringerLink, and IEEE Xplore}.
To further improve the comprehensiveness of our search and identify relevant literature not published in indexed journals, such as preprints, a complementary search was conducted in \textbf{Google Scholar}.

A unified search query was designed to strike a balance between precision and recall, directly informed by our RQs. The query is structured around three core concepts connected by the `AND` operator: (1) the AI methodology, (2) the problem domain, and (3) the object of study. The final logical query was:

\begin{center}
\textit{( "Deep Learning" OR "Neural Network" OR "Neural Networks" ) AND ( "Presentation Attack Detection" OR "PAD" OR "anti-spoofing" OR "document forgery" OR "identity fraud" ) AND ( "identity document" OR "identity documents" OR "ID cards" OR "ID card" OR "identity cards")}
\end{center}

This query was adapted to the specific syntax of each database. The search was limited to articles published between January 2020 and July 2025. Table \ref{tab:query_search_results} summarises the databases searched and the number of initial hits obtained from each.

\begin{table*}[]
\centering
\caption{Primary databases and initial search results (July 2025).}
\label{tab:query_search_results}
\begin{tabular}{@{}lc@{}}
\toprule
\textbf{Database} & \textbf{Nº Publications Found} \\ \midrule
Scopus & 97 \\
ACM Digital Library & 131 \\
SpringerLink & 77 \\
IEEE Xplore & 130 \\
\midrule
\textbf{Subtotal from Primary Databases} & \textbf{435} \\
\midrule
Google Scholar (Supplementary Search) & 892 \\
\bottomrule
\end{tabular}
\end{table*}

\subsection{Inclusion and Exclusion Criteria}
To systematically filter the retrieved articles, a set of precise inclusion and exclusion criteria was defined, aligned with the scope of our RQs.

\textbf{Inclusion Criteria:}
\begin{itemize}
\item Primary research articles (from journals or conferences) published between 2020 and 2025.
\item Articles written in English.
 \item Studies that propose, evaluate, or analyse a Deep Learning/Foundation model-based method for:
\begin{itemize}
 \item a) The detection of presentation attacks in a physical identity document verification process.
 \item b) The generation of synthetic data for the aforementioned purpose.
\end{itemize}
 \item Studies that introduce or analyse a dataset specifically created for PAD on identity documents.
\end{itemize}

\textbf{Exclusion Criteria:}
\begin{itemize}
 \item Duplicate articles found across multiple databases.
 \item Review articles, editorials, patents, or book chapters.
\item Studies focused exclusively on other biometric modalities (e.g., PAD in a face recognition system without document analysis, fingerprint, iris) or on non-biometric fraud (e.g., financial fraud).
\item Articles unavailable in full-text format.
\end{itemize}

\subsection{Study Selection Process}
\begin{figure*}[]
\centering
\includegraphics[width=0.8\linewidth]{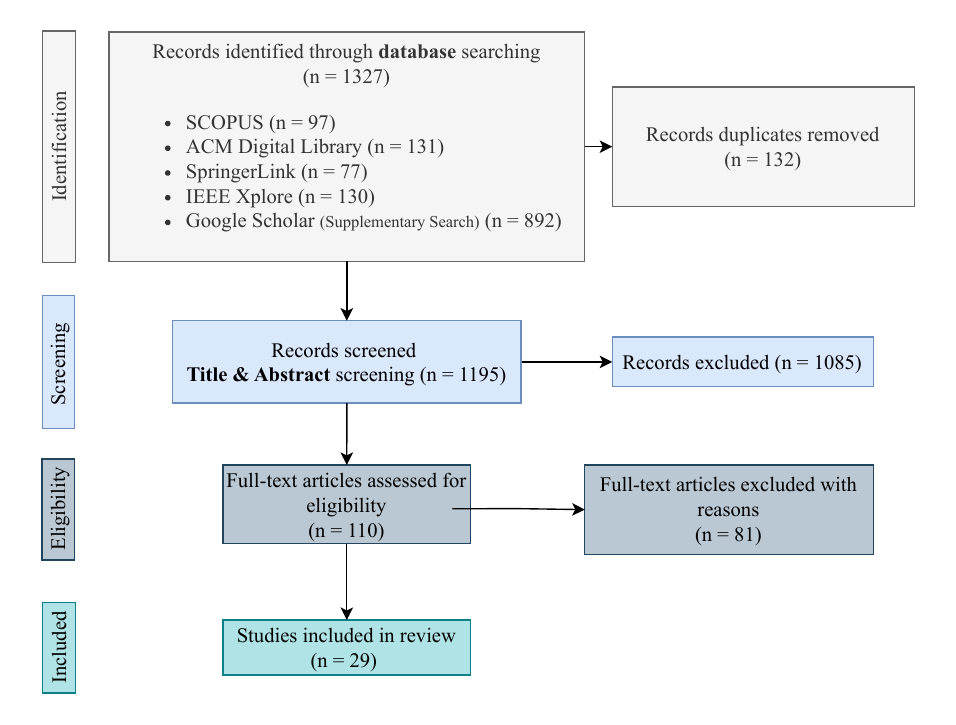}
\caption{PRISMA flow diagram illustrating the study selection process.}
\label{fig:prisma_flowchart}
\end{figure*}

The study selection process, illustrated in the PRISMA flowchart in Figure \ref{fig:prisma_flowchart}, was conducted in three phases.

Initially, the database search yielded a total of \textbf{1327} records (435 from primary databases and 892 from Google Scholar). In the first phase, \textbf{132} duplicate records were identified and removed, leaving 1195 unique articles. In the second phase, these articles underwent a title and abstract screening based on the inclusion/exclusion criteria. This led to the exclusion of \textbf{1085} articles that were clearly out of scope (e.g., focused merely on biometric modalities or applications). The remaining \textbf{110} articles were sought for full-text retrieval and assessed for eligibility in the third phase. During this detailed review, another \textbf{81} articles were excluded for reasons such as the work was not specifically focusing on identity document PAD or other inclusion/exclusion criteria.

\begin{figure*}[]
 \centering
 \includegraphics[width=0.6\linewidth]{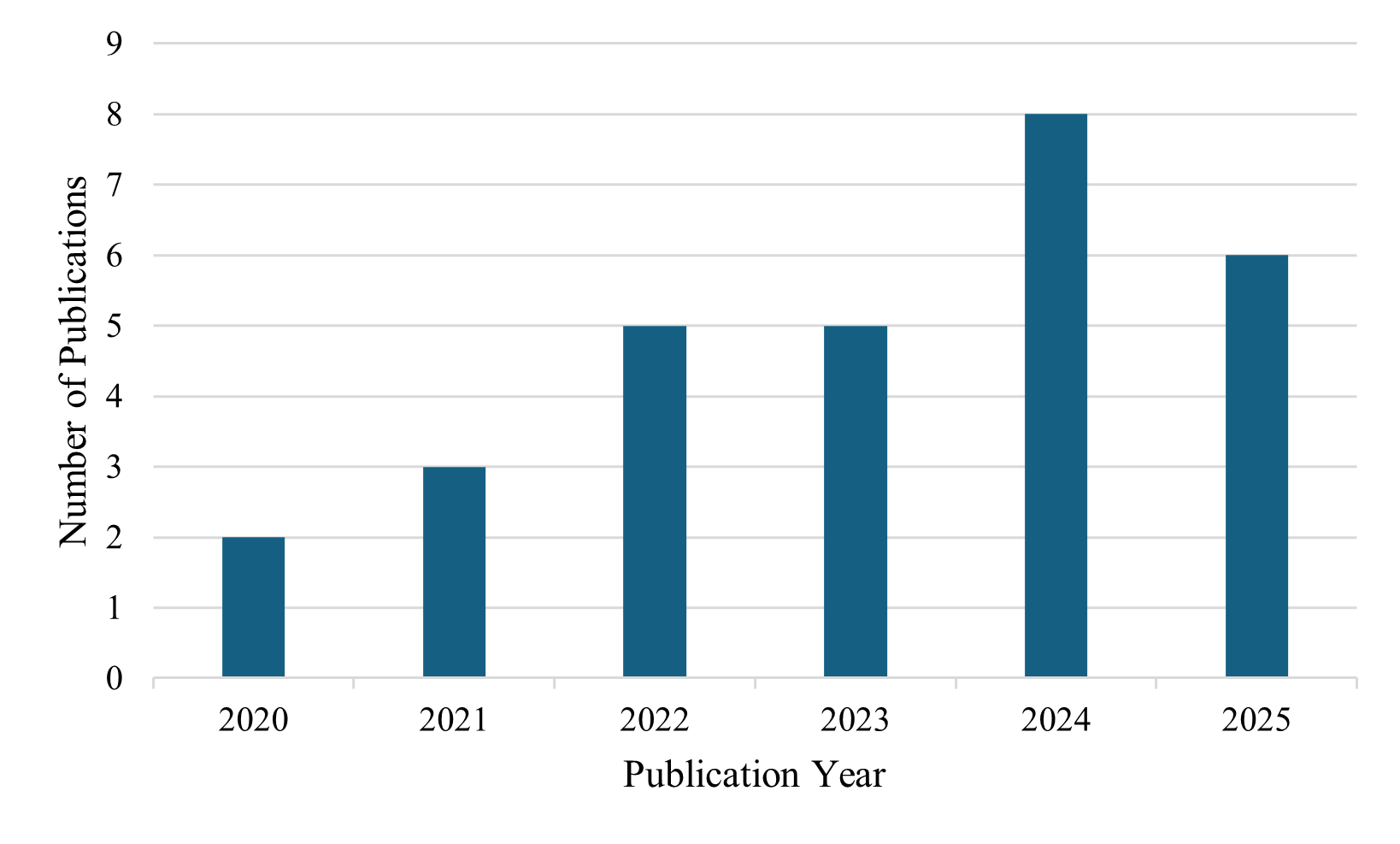}
 \caption{Distribution of selected primary studies on PAD for identity documents by publication year. The chart illustrates the sustained and growing research interest in the field within our review period (2020-2025).}
 \label{fig:publication_distribution}
\end{figure*}

This process resulted in a final selection of \textbf{29} primary studies considered highly relevant to our research questions. These include \textbf{16} studies focusing on novel methodologies for Presentation Attack Detection and synthetic data generation (addressing RQ2 and RQ4), and \textbf{13} key papers that introduce or analyse the most widely used public datasets in the field (addressing RQ1). The distribution of these selected articles by publication year, as depicted in Figure \ref{fig:publication_distribution}, clearly illustrates a sustained and growing research interest in PAD for identity documents within the reviewed period, especially noting the recent increase in publications.

To evaluate the reliability and alignment of the included studies, a quality assessment (QA) was performed on the 16 methodological papers. Each study was rated against four key quality criteria (QA1-QA4):
\begin{itemize}
\item \textbf{QA1: Clarity of Objectives:} Are the research goals and problem statement clearly defined?
 \item \textbf{QA2: Methodological Rigour:} Is the proposed DL/FM architecture and experimental setup described with sufficient detail to allow for conceptual replication?
 \item \textbf{QA3: Dataset Transparency:} Is the dataset used for evaluation adequately described (size, composition, bona fide/attack types)? Is it public or private?
 \item \textbf{QA4: Evaluation Validity:} Are standardised performance metrics (e.g., APCER, BPCER, EER) used? Are the conclusions well-supported by the results?
\end{itemize}

Overall, the selected studies demonstrated high quality, with all papers clearly stating their objectives and using valid evaluation metrics. However, a significant finding from the QA process was the lack of transparency and accessibility in the dataset. We found that over 60\% of the methodological studies relied on private or in-house datasets, a critical barrier to reproducibility and comparative benchmarking in the field. This finding reinforces the importance of RQ1 and highlights the need for more high-quality, publicly available benchmark datasets.

\subsection{Data Extraction}
A standardised data extraction form was developed to systematically collect relevant information from the 29 selected articles. The extracted data points were directly mapped to the Research Questions defined in Section \ref{subsec:rqs}:
\begin{itemize}
\item \textbf{For RQ1 (Datasets):} We extracted dataset names, year of publication, scope (document types), nature of bona fide samples (e.g., mock-ups, synthetic templates), and the specific attack types modelled.
\item \textbf{For RQ2 (Architectures):} We documented the core DL/FM architectures (e.g., MobileNet, DenseNet, VisonTransformers), training paradigms (e.g., two-stage, few shot learning, adversarial), and key technical innovations.
 \item \textbf{For RQ3 (Performance):} We collected the reported performance metrics, primarily APCER, BPCER, and EER, along with the datasets on which these results were achieved.
 \item \textbf{For RQ4 (Synthetic Data \& Gaps):} We recorded the generation methods (e.g., GANs, texture transfer), the models used (e.g., CycleGAN, StyleGAN), and the metrics used to evaluate the quality of synthetic data (e.g., FID), as well as synthesized the primary challenges discussed.
\end{itemize}
This structured extraction process enabled a consistent and comprehensive synthesis of the findings, which are presented in the subsequent sections of this review.

\subsection{Differences from Previous Studies}

Compared to previous research on presentation attacks in biometric recognition systems, such as facial recognition \cite{Raghavendra-FacePAD-Survey-2017,  abdullakutty_review_2021}, fingerprint \cite{Sousedik-PAD-Survey-IET-BMT-2014,ametefe_advancements_2025}, or iris recognition \cite{agarwal_presentation_2021}, the study of attack detection in identity document verification processes remains a rather nascent field. 

While several reviews have addressed fraud detection across various biometric modalities or specific anti-spoofing techniques \cite{shaheed_deep_2024}, to the best of our knowledge, a dedicated systematic review that comprehensively analyses and synthesises recent advancements in DL, FMs methodologies, and synthetic data generation specifically for PAD targeting identity documents has not yet been presented. 

Existing literature often provides a broader overview of biometric security or focuses on individual PAD techniques without a concentrated analysis of the unique challenges, datasets, and evolving solutions pertinent to the verification of physical identity credentials. This review aims to fill this identified gap by providing a detailed and focused synthesis of recent developments in this specialised domain. We establish a consolidated reference framework for future research by systematically examining computational architectures, the landscape of available datasets (including their limitations), evaluation metrics, and data augmentation strategies, thereby highlighting distinct challenges such as document variability, the intricacies of Presentation Attack Instrument (PAI) creation for diverse document types, and the critical need for improved generalisation.

\section{Overview}
\label{sec:pad_methods}
\subsection{Deep Learning in Presentation Attack Detection (PAD)}

In recent years, DL has fundamentally reshaped the landscape of PAD for identity documents \cite{shaheed_deep_2024, abdullakutty_review_2021}. Its strength lies in the ability of models, especially various Convolutional Neural Network (CNN) architectures—ranging from lightweight networks optimized for mobile deployment (like MobileNet variants \cite{sandler_mobilenetv2_2019}) to deeper models focused on robust feature extraction (such as ResNet variants like ResNeXt \cite{xie_aggregated_2017} or DenseNet families \cite{huang_densely_2017}) — to autonomously learn highly discriminative representations directly from document images. These models excel at identifying subtle inconsistencies and artefacts that often betray manipulation attempts, such as anomalies in texture patterns \cite{magee_investigation_2024}, unnatural edges from composite images (splicing), illumination inconsistencies, specific printing artefacts (e.g., dot patterns, banding), or visual cues from screen recaptures (like pixel grids and moiré effects) \cite{mudgalgundurao_pixel-wise_2022}.

Current research explores a spectrum of DL strategies: end-to-end classifiers distinguishing bona fide from attack presentations \cite{gonzalez_forged_2025, gonzalez_towards_2022, gonzalez_hybrid_2021}, methods employing pixel-level supervision to precisely locate forged regions \cite{mudgalgundurao_pixel-wise_2022}, Siamese or triplet networks designed for comparative analysis between questioned documents and known references or templates \cite{chen_domain-agnostic_2022}, and few-shot learning approaches aiming to rapidly adapt detection capabilities to novel or sparsely represented attack vectors \cite{sanchez_few-shot_2024}. 

Despite these advancements, the practical efficacy of DL in PAD on ID-cards remains significantly challenged by its reliance on extensive, diverse, and well-balanced training datasets. The inherent abundance of real-world attack samples compared to the scarcity of bona fide documents (data imbalance) \cite{wang_review_2021, tyagi_sampling_2020} critically limits the generalisation performance of trained models, potentially leading to vulnerability against unseen attack types. This limitation becomes particularly pronounced when attempting to develop models that can generalise effectively across diverse identity documents from multiple countries or issuers, especially when training data is predominantly sourced from a limited set of document types. Addressing this cross-domain generalisation challenge, often through techniques such as few-shot learning or zero-shot learning, is thus a critical area of ongoing research.

Furthermore, the absence of universally accepted benchmarking datasets and standardised evaluation protocols \cite{ISO-IEC-30107-3-PAD-metrics-2023} makes rigorous comparison between proposed methods difficult and hinders the transition of research innovations into reliable operational systems.

\subsection{Hybrid and Multi-Stage Approaches for Specialised Detection}

A predominant trend in the PAD field is the development of multi-stage systems that decompose the detection problem into more specific tasks, assigning specialised neural networks to each. This approach enables the optimisation of detecting different attack families, such as digital manipulations (composition attacks) and physical attacks (printouts and screen captures).

The works of Gonzalez et al. \cite{gonzalez_hybrid_2021} illustrate a clear evolution of this philosophy. Initially, they propose a hybrid two-stage system for chipless Chilean identity documents. The first stage uses a MobileNet, pre-trained on ImageNet and fine-tuned through transfer learning, to distinguish between "composite" (digitally manipulated using \textit{splicing} techniques) and "non-composite" images. The images classified as "non-composite" proceed to a second stage, where a lightweight convolutional network named BasicNet, trained from scratch, classifies them as "real," "printed," or "screen-displayed." The key innovation of BasicNet lies in the pre-processing of the input, demonstrating that the Discrete Fourier Transform (DFT) is highly effective for extracting high-frequency features that differentiate the unique textures of real cards from print patterns or screen pixels.

Subsequently, this architecture is refined and evaluated in \cite{gonzalez_towards_2022}, where the two-MobileNetV2 system is maintained, but a critical variable is introduced: the Face Image Quality Assessment (FIQA) index. Using the MagFace method \cite{meng_magface_2021} to evaluate the quality of the photograph on the document, the study analyses how low-quality images impact the system's ability to detect attacks. The results suggest that prior filtering based on the quality of the entire document image can improve system performance, although its impact is reduced when analysing only the photograph section, indicating a non-trivial correlation between the quality of the face image and the quality of the entire document.

The culmination of this approach is presented in \cite{gonzalez_forged_2025}, where the two-stage architecture with MobileNetV2 is trained from scratch on an extensive private dataset of $190,000$ images, eliminating the dependency on ImageNet and adapting to multiple Chilean card formats (CHL1 and CHL2). This work significantly expands the attack classes, including not only print, screen, and composition attacks, but also synthetic (digitally generated) and PVC card attacks. A fundamental contribution is the development of PyPAD\footnote{\url{https://github.com/jedota/PyPAD}}, an evaluation framework compliant with the ISO/IEC 30107-3 standard \cite{ISO-IEC-30107-3-PAD-metrics-2023} that allows for rigorous multi-class analysis. The concatenation of the two networks, along with prior semantic segmentation to remove the image background, demonstrates unprecedented accuracy, establishing a robust and specialised system for remote verification contexts.

\subsection{Forensic Analysis Based on Microscale Artefacts and Textures}

Another line of research focuses on detecting the subtle and unavoidable artefacts that the printing, scanning, and recapture processes leave on the ID-card image. These methods, operating at the pixel or patch level, search for forensic traces such as moiré patterns, compression artefacts, paper textures, and colour distortions.

An example of a classic-modern approach is that of Magee et al. \cite{magee_investigation_2024}, who investigate the use of the Meijering filter \cite{meijering_design_2004}, a biomedical image processing algorithm, for recapture detection. By applying the filter to grayscale images, they extract textural features that magnify the differences between documents recaptured from a screen and those printed on paper. Instead of a deep neural network, they use statistical tests (Chi-Square, t-Student) on intensity histograms and a Support Vector Machine (SVM) classifier to distinguish between the two classes. Their results validate that a specialized filter, even without a deep learning feature extractor, can be effective in detecting the textural differences inherent to different types of presentation attacks.

In contrast, \cite{mudgalgundurao_pixel-wise_2022} addresses the same problem from a deep learning perspective with pixel-wise supervision. They propose an architecture based on DenseNet \cite{xie_aggregated_2017}, chosen for its ability to optimize information flow in deep networks, and combine it with dual supervision: a binary classification at the image level and pixel-wise supervision. This allows the network to learn to detect low-level patterns, such as moiré effects or print artefacts, without the need for explicit pre-processing. By forcing the network to pay attention to every region of the image, the detection of subtle manipulations that might go unnoticed in a global classification scheme is improved.

Similarly, \cite{chen_domain-agnostic_2022} adopts a forensic comparison approach using a Siamese network. Instead of classifying an image in isolation, the model processes triplets of image patches: one from a genuine document, one from a high-quality reference, and one from a recaptured document. Using a ResNeXt101-based backbone \cite{xie_aggregated_2017} and a forensic similarity subnetwork, the system learns an embedding space where genuine patches cluster near the reference, while recaptured ones are pushed away. The combination of Triplet Similarity Loss and Normalised Softmax Loss optimises this feature space, preserving the "forensic signature" of each document type and improving generalisation against domain variations such as different capture devices or printing substrates.

Taking forensic analysis a step further, Chen et al. \cite{chen_multi-modal_2024} propose a method that not only identifies forensic traces but explicitly \textbf{disentangles} them for use as inputs to a multimodal model. Their work, designated MMDT (Multi-modal Disentangled Traces), addresses the limited generalisation faced by other methods that rely on manual data collection or knowledge of device parameters \cite{chen_distortion_2024, benalcazar_synthetic_2023}. The architecture first employs a self-supervised disentanglement and synthesis network to extract the "recaptured traces" from an image. These traces are defined as two distinct components: the \textbf{blur content (C)} and the \textbf{texture components (T)}, corresponding to the blurring and halftoning distortions, respectively. The fundamental innovation of this approach is treating these disentangled forensic traces as \textbf{new modalities} of data, rather than relying solely on the RGB domain. Subsequently, a transformer (specifically, ViT-B16) ingests all three modalities (RGB, C, and T), efficiently fusing the information via lightweight \textbf{Adaptive Multi-modal Adapters (AMA)}. By explicitly employing the distortion traces as a multimodal input, the model significantly enhances its generalisation capability in cross-domain scenarios, without requiring the manual effort of collecting additional data or prior knowledge of acquisition devices.

Following a similar logic of fusing forensic artefacts, Li et al. \cite{li_two-branch_2023} propose a two-branch, multi-scale architecture to improve generalisation in recapture detection. Their method processes two input modalities in parallel: one branch uses a \textbf{frequency filter bank} (Filter Bank Process) based on DCT to extract \textbf{fine detail loss} artefacts, while the second branch operates on the \textbf{original RGB image} to capture \textbf{colour distortion}. Instead of a simple concatenation, a \textbf{multi-scale cross-attention fusion} module aligns and combines these two forensic traces at different levels of the network. This explicit fusion of textural and colour artefacts demonstrates significantly improved detector robustness in cross-dataset and cross-quality evaluation scenarios.

From an offensive perspective, which is fundamental to understanding the vulnerabilities that PAD systems must address, the work by Zhao et al.~\cite{Zhao} introduces a comprehensive document forgery framework, named \textbf{ForgeNet}, which specialises in realistic text editing, even on documents with complex characters (such as Chinese) and elaborate background textures. The relevance of this work to forensic analysis lies not only in its text editing capabilities but also in its sophisticated post-processing approach, explicitly designed to conceal microscale artefacts. The ForgeNet framework is composed of three key modules: a text editing network (TENet) that improves upon existing methods; a pre-compensation network (\textbf{PCNet}) that adds colour and noise distortions typical of the print-and-scan channel to the edited region to ensure forensic consistency with the rest of the document; and an inverse halftoning network (\textbf{IHNet}) that removes existing dot patterns before the physical recapture process. This final stage is crucial, as it prevents interference between the halftone patterns of the forged document and those that would be generated during recapture, thereby avoiding the creation of tell-tale artefacts such as Moir{é} patterns. The most alarming contribution of this work is the demonstration that the forged-and-recaptured documents created with their method successfully \textbf{fooled several state-of-the-art document authentication systems}. This study sets a new benchmark for the sophistication of attacks, highlighting the need for forensic detection methods to evolve beyond the analysis of obvious artefacts to identify manipulations that have been deliberately engineered to be forensically consistent.

\subsection{Synthetic Data Generation for Data Imbalance Mitigation}

As a direct response to the critical data scarcity, compounded by the aforementioned privacy constraints that restrict the broad dissemination of bona fide document images, and imbalance challenges \cite{wang_review_2021} hindering Deep Learning-based PAD systems, synthetic data generation has gained significant traction within the research community \cite{benalcazar_synthetic_2023, markham_open-set_2024}. 

The core motivation is to automatically create synthetic, realistic bona fide samples as well as Presentation Attack samples, thereby reducing the costs, demanding time, privacy concerns, and logistical difficulties associated with collecting large volumes of diverse, real-world attack data. 

Generative Adversarial Networks (GANs) \cite{ye_generative_2022}, have been used in several applications and architectures such as CycleGAN \cite{chu_cyclegan_2017} (suited for unpaired image-to-image translation), pix2pix \cite{pix2pix2017}(for paired translation), and others \cite{markham_open-set_2024}, stand out as prominent tools for this task. Researchers employ these models to synthesise visual artefacts characteristic of specific PA types, aiming to replicate the nuances of printed documents (simulating different paper substrates and printer technologies) and screen recaptures (emulating display characteristics and camera interactions), as explored in \cite{benalcazar_synthetic_2023, markham_open-set_2024}. Beyond GANs, explorations may include procedural generation using computer vision algorithms or advanced texture transfer algorithms \cite{benalcazar_synthetic_2023, li_sitta_2023}. 

The overarching goal is to enrich limited real training datasets with diverse, synthetically generated PAIs, thereby improving the robustness and generalisation capacity of DL detection models against a wider range of attack scenarios. However, the utility of synthetic data is not without its own set of challenges. 

Ensuring that generated samples possess sufficient fidelity and accurately reflect the variability of real bona fide samples and real attack samples remains a complex issue, often requiring sophisticated validation techniques and metrics (like the Fréchet Inception Distance - FID \cite{jayasumana_rethinking_2024}, used in \cite{benalcazar_synthetic_2023, markham_open-set_2024}) to quantify realism. 

Moreover, there's a persistent risk that models might overfit to artefacts specific to the synthetic generation process rather than learning genuine attack cues. Developing standardised methodologies for generating and validating synthetic data, and carefully integrating it into training regimes to avoid bias, are crucial ongoing research efforts \cite{benalcazar_synthetic_2023}.

It is essential to highlight that there have been only a few attempts to measure or develop metrics specifically to evaluate the synthetic generation or the quality of the entire ID-card. Most attempts in this area are more closely aligned with face image quality assessment, such as OFIQ \cite{Merkle-OFIQ-Report-240930, bsi201PublicReportV1_2}.

\subsection{Advanced Learning Paradigms for Generalisation and Data Scarcity}

One of the greatest challenges in ID-Card PAD is the scarcity of public data and the difficulty in obtaining samples of genuine identity documents from multiple countries and various types of attacks. To overcome these limitations, research has shifted towards more advanced learning paradigms which seek to improve the generalisation capabilities of models, either by learning from a few examples or by synthetically augmenting the available data. These strategies can be broadly categorised into three main approaches: making models more data-efficient, making training more robust, and creating new data.

The first paradigm, Few-Shot Learning (FSL), directly addresses the problem of extending attack detection to new, unseen document types with minimal data. The study developed by \cite{sanchez_few-shot_2024} employs prototypical networks, a metric-based FSL approach, to classify screen recapture attacks. Starting with a base model trained on documents from Spain and Chile, they demonstrate effective generalisation to documents from Argentina and Costa Rica by adding only a small "support set" from the new country. This methodology drastically reduces the need to collect large volumes of sensitive data, opening a promising path for the rapid adaptation of PAD systems to new markets.

A second paradigm involves adversarial learning to improve robustness against unknown variations. The work in \cite{chen_distortion_2024} focuses on enhancing recapture detection in cross-domain scenarios through an adversarial substitution model. A CycleGAN-based model is used to generate realistic recapture distortions, which is then integrated into a competitive training loop with the detection model (DPAD). The generator constantly creates new synthetic attacks to generalisation detector, forcing the detector to learn more robust and generalizable features that surpass the metrics of previous methods in cross-domain settings.

Following a similar line of research on cross-domain robustness, but replacing adversarial learning with a model-based forensic approach, Chen et al. \cite{chen_spectral_2024} address the same challenge of recapture detection. Their work is predicated on the premise that existing frequency-domain augmentation (FDA) methods are suboptimal because they treat all spectral bands equally. To address this, they propose a two-phase method. First, they establish a Band-of-Interest Localisation (BOIL) technique, which utilises domain knowledge and the generalisation of models (derived from printing and imaging processes) to precisely identify the specific frequency bands (the 'Band-of-Interest' or BOI) where the forensic artefacts of recapture, such as halftoning patterns, reside. Secondly, they introduce a Frequency-domain Halftoning Augmentation (FHAG) strategy, which applies tailored augmentation operations (such as halftone mixing and noise addition) exclusively within this identified BOI. By focusing the augmentation solely on the relevant forensic features, they improve the model's robustness without distorting other semantic information. To validate their method, they also constructed a new diverse dataset, RDID162, demonstrating a significant reduction in Equal Error Rates (EER) in cross-domain scenarios compared to other FDA strategies.

A third, complementary strategy is the synthetic generation of attack samples to directly tackle data scarcity and imbalance. Benalcazar et al. \cite{benalcazar_synthetic_2023} conducted a foundational exploration into this domain, assessing multiple generation methods including computer vision algorithms, texture transfer techniques, and Generative Adversarial Networks (GANs)\footnote{\url{https://github.com/jedota/Synthetic_ID-Card_Image}}. Their work established the viability of augmenting training sets with synthetic data, showing that this did not significantly degrade detector performance and that techniques like texture transfer and CycleGAN were highly effective for creating realistic print attack samples. 

Building on this, Markham et al. \cite{markham_open-set_2024} provided a focused comparative study of different GAN architectures (pix2pix, CycleGAN, etc.) using open-source datasets, making their findings more reproducible. They found that unsupervised models like CycleGAN produce visually faithful samples that can improve or maintain PAD performance. A key insight from their work is the potential disconnect between visual fidelity (measured by Fréchet Inception Distance - FID) and the utility of the data for training, suggesting the accurate replication of specific attack artefacts is more critical than overall visual perfection. Collectively, these studies establish synthetic data generation as a validated methodology for automating PAI creation, enhancing model robustness, and enabling privacy-preserving research.

\subsection{Foundation models}

In response to the persistent challenges of data scarcity and poor cross-domain generalisation, the research community is turning towards \textbf{Foundation Models (FMs)}, a class of models that represent a true paradigm shift. Unlike traditional DL approaches that learn features from scratch for a specific task, FMs leverage knowledge pre-trained on web-scale datasets, fundamentally changing the approach from task-specific learning to knowledge adaptation via transfer learning.

This category encompasses a diverse range of architectures, from powerful vision models like DinoV2 \cite{Dinov2} to multimodal vision-language models, such as Contrastive Language-Image Pre-Training (CLIP) \cite{CLIP}, which serve as highly effective feature extractors. Their capabilities can be harnessed through various strategies, such as direct application in zero-shot scenarios \cite{sanchez_few-shot_2024} or adaptation through fine-tuning a small part of the network \cite{tapia_can_2025}. A notable example of the latter is the use of FMs within novel, privacy-preserving frameworks, where models like DinoV2 are trained on small, anonymised image patches rather than full documents, demonstrating high efficacy in detecting forensic traces while minimising data exposure \cite{FakeIDet}.

By capitalising on these robust, pre-existing visual representations, FMs offer a path to create highly generalisable PAD systems without the need for millions of domain-specific images. This potential to overcome the critical data bottleneck marks them as a crucial direction for future research, whose specific implementations and performance are analysed in detail in Section \ref{sec:pad_methods}.

Presentation Attack Detection (PAD) is a fundamental pillar for the security of remote identity verification systems. Faced with the growing sophistication of document fraud, the scientific community has transitioned from monolithic approaches to specialised and robust solutions. This section analyses recent advances in PAD methods, organising them according to the key methodological trends that define the current state of the art: the use of hybrid and multi-stage architectures, the forensic analysis of microscale artefacts, and the application of advanced learning paradigms to improve generalisation and address data scarcity.

\subsection{Leveraging Foundation Models for PAD Generalisation}

The formidable challenge of achieving broad generalisation capabilities across diverse ID documents and attack types, particularly when faced with limited training data, has led the research community to explore advanced paradigms. Among these, the application of \textbf{Foundation Models (FMs)}, pre-trained on vast and varied datasets, presents a highly promising avenue for enhancing Presentation Attack Detection (PAD) systems on ID-cards.  Unlike traditional deep learning models that often struggle to generalise to unseen data due to privacy concerns limiting dataset size, FMs offer an inherent capacity for transfer learning, acting as robust feature extractors.

Tapia and Busch \cite{tapia_can_2025} conducted a seminal study investigating the generalisation capabilities of FMs, specifically \textbf{DinoV2} and \textbf{CLIP}, for ID-card PAD. Their work explored different test protocols, including zero-shot and fine-tuning approaches, utilising both a private Chilean ID dataset and an open-set dataset comprising ID-cards from Finland, Spain, and Slovakia.

Their findings highlight several critical insights:
\begin{itemize}
    \item \textbf{Superior Generalisation:} FMs, particularly DinoV2, consistently outperformed traditional deep learning models and vision transformers (like Swin-Transformer) in various generalisation scenarios.  This superior performance is attributed to the enormous quantity and diversity of data on which these FMs were initially pre-trained, even if that data did not explicitly include ID-cards.
    \item \textbf{Role of Bona Fide Images:} A crucial discovery was that the quality and representativeness of \textbf{bona fide} images are key to achieving strong generalisation capabilities. The study indicated that models perform significantly better when trained with genuine images that accurately reflect real-world characteristics, rather than relying solely on a large volume of attack samples.  This is particularly relevant for datasets like ID-Net, where simulated bona fide images are synthetically generated from templates, potentially confusing classifiers due to their similarity with generated attacks.
    \item \textbf{Zero-Shot and Fine-Tuning Performance:} In zero-shot scenarios, DinoV2-vitl14 achieved an Equal Error Rate (EER) of 4.33\% for border attacks, demonstrating its capacity to generalise without explicit training on new attack types.  When fine-tuned with a small classification head, DinoV2-vis114 further improved performance, reaching EERs of 4.13\% for composite/border attacks, 1.28\% for printed attacks, and 2.73\% for screen attacks, showcasing the benefits of adapting these powerful models to specific PAD tasks.
     \item \textbf{Fusion for Enhanced Performance:} To further boost performance, especially on challenging datasets like ID-Net, Tapia and Busch proposed a fusion approach at the score level, combining the strengths of FMs (excelling at local feature extraction) with traditional deep learning models (better for global features).  This hybrid method significantly reduced the EER on ID-Net from 27.86\% (for the best individual model) to 8.25\%, demonstrating the synergistic benefits of leveraging complementary feature representations.
\end{itemize}
This pioneering work underscores that Foundation Models represent a paradigm shift for ID-card PAD, offering a pathway to overcome data scarcity and achieve robust generalisation, even for unknown attack types and documents from diverse countries.  The emphasis on bona fide image quality and the potential of multi-model fusion are critical lessons for future research.

Complementing this line of research, Mu{ñ}oz-Haro et al. \cite{FakeIDet} introduced \textbf{FakeIDet}, a framework that leverages FMs not only to improve generalisation but also to pioneer a novel \textbf{privacy-preserving methodology}. Their work directly confronts the critical barrier of data privacy, which has historically prevented the creation of public datasets containing samples from genuine ID-cards. Instead of processing entire ID images, FakeIDet employs a two-stage process: first, an ID document is fully or pseudo-anonymised by obscuring sensitive fields; subsequently, the system extracts a set of small, non-identifiable image patches (e.g., $64\times64$ pixels) from non-sensitive areas. These patches are then fed into a fake patch detector, for which the study demonstrates the superior performance of a \textbf{DinoV2} model with a fine-tuned classification head. This patch-wise approach proved remarkably effective, achieving 0\% EER at the full ID level on their internal test set, starkly outperforming the 33.3\% EER obtained when applying the same model to full ID images. Crucially, the model demonstrated outstanding generalisation in a \textbf{cross-database scenario}, maintaining at the ID-card level when tested on the unseen DLC-2021 dataset.

\begin{table*}[]
\scriptsize
\centering
\caption{Articles found by the systematic literature review}
\label{tab:articles_found_detection}
\begin{tabularx}{\textwidth}{@{}lXXl@{}}
\toprule
Nº Art & Title & Algorithms  & Datasets \\ \midrule

\cite{li_two-branch_2023} & Two-Branch Multi-Scale Deep Neural Network for Generalised Document Recapture Attack Detection & ResNet50  & Private used for \cite{chen_domain-agnostic_2022} \\

\cite{chen_multi-modal_2024} & Multi-Modal Document Presentation Attack Detection with Forensics Trace Disentanglement
 & ViT-B16  & RSCID, RDID162 and RSCID(L) \\

\cite{chen_spectral_2024} & Distortion Model-Based Spectral Augmentation for Generalized Recaptured Document Detection & FHAG \& BOIL + ResNet50, DenseNet121   & RDID162 \\

\cite{FakeIDet} & FakeIDet: Exploring Patches for Privacy-Preserving Fake ID Detection & Foundation Model (DINOv2), Vision Transformer (ViT-B/16), ResNet-18, & FakeIDet-db \\

\cite{tapia_can_2025} & {Can Foundation Models Generalise the Presentation Attack Detection Capabilities on ID-cards?} & Foundation Models (DinoV2-small, DinoV2-base, DinoV2-large; CLIP-VIT-B16, CLIP-VIT-B32, CLIP-VIT-L14) & Private Chilean, ID-Net \\

\cite{chen_distortion_2024}   & A distortion model guided adversarial surrogate for recaptured document detection  & CycleGAN and ConvNeXtTiny  & The Recaptured Student RSCID \cite{chen_domain-agnostic_2022} \\

\cite{magee_investigation_2024}  & An Investigation into the Application of the Meijering Filter for Document Recapture Detection & Histogram and SVM classification & BID \cite{soares}                \\

\cite{sanchez_few-shot_2024}  & Few-Shot Learning: Expanding ID-cards Presentation Attack Detection to Unknown ID Countries     & EfficientNetV2-B0   & Private DB                \\

\cite{gonzalez_forged_2025} & Forged presentation attack detection for ID-cards on remote verification systems                & MobileNetV2  & Private DB                \\

\cite{markham_open-set_2024} & Open-Set: ID-card Presentation Attack Detection using Neural Transfer Style & Pix2pix and pix2pixHD, U-Net, MobiletV2 & MIDV2020\cite{bulatov_midv-2020_2022}, DLC2021 \cite{v_document_2022}              \\

\cite{benalcazar_synthetic_2023}  & Synthetic ID-card Image Generation for Improving Presentation Attack Detection & StyleGAN2-ADA, CycleGAN
MobileNetV2  & Semi-Private \\

\cite{mudgalgundurao_pixel-wise_2022}     & Pixel-wise supervision for presentation attack detection on identity document cards & CNN & Private                \\

\cite{chen_domain-agnostic_2022} & Domain-Agnostic Document Authentication Against Practical Recapturing Attacks                & ResNet101 & Private                \\

\cite{gonzalez_towards_2022}  & Towards Refining ID-cards Presentation Attack Detection Systems using Face Quality Index       & Hybrid two-stages MobileNetv2  & Private  \\

\cite{gonzalez_hybrid_2021} & Hybrid Two-Stage Architecture for Tampering Detection of Chipless ID-cards & MobileNet + DFT filter & Private \\

\cite{Zhao} & Deep Learning-Based Forgery Attack on Document Images & CNN ForgeNet  & Student Card Chinese DB \\
\\ 
\bottomrule
\end{tabularx}
\end{table*}

Beyond the methodology, the work's most significant contribution is the release of \textbf{FakeIDet-db}, a dataset comprising patches from 20 \textbf{bona fide identity documents}, which addresses the critical data availability issue that has long hindered reproducible research in the field. This work, therefore, establishes a viable pathway for training highly effective and generalizable PAD systems on data derived from authentic documents without compromising individual privacy.  As a summary of what was previously exposed by each research, see the \textbf{Table \ref{tab:articles_found_detection}}. 

\subsection{Benchmarking in Academic Competitions}

Beyond the proposal of individual models, academic competitions provide a benchmark of different computational models developed and designed by various engineering teams, evaluating multiple approaches under a standardised and unbiased framework. 

Tapia et al. \cite{tapia_first_2024}, \cite{tapia_second_2025} have raised the issue of creating a third-party evaluation process to remove the bias developed by each algorithm, which is trained and tested using a similar protocol to generate fake ID-cards. The first Presentation Attack Detection for Identity Cards (PAD-ID-card) competition held at the International Joint Conference on Biometrics (IJCB) 2024 provides a valuable snapshot of the field's current capabilities and persistent challenges. The best team achieved an EER of 21.87\% and an Average rank of 74,30\%. The second edition of this competition was also organised at IJCB in 2025, considering two tracks in order to evaluate the performance of the algorithm on private and open-set datasets. The best team for Track 1 achieved an EER of 11.34\% and an Average rank of 40,48\%. For Track 2 the best results was an EER of 6,36\% and an Average rank of 14,76\% \footnote{https://sites.google.com/view/ijcb-pad-id-card-2025/}.

The competition provided an independent assessment by evaluating eight distinct models from five teams on a sequestered test dataset, which included ID-cards from four different countries and various attack types (composite, print, and screen). A key finding of the competition was the significant difficulty that state-of-the-art models face with \textbf{cross-dataset generalisation}. The results revealed a notable performance gap between models trained on private datasets versus those trained on publicly available ones, underscoring the limitations of current open-source data.

Furthermore, the competition highlighted that screen replay attacks remain one of the most challenging PAIs to detect consistently. The study also observed that models performed better on ICAO-compliant documents, suggesting that standardisation in document design aids in the generalisation of detecting attacks. By establishing a formal evaluation protocol and public leaderboard, this work not only benchmarks current methods but also provides a vital framework for future research to measure progress against a common, challenging standard.

Very recently, Korshunov et al. proposed a new challenge at the International Conference on Computer Vision (ICCV-2025) called "The Challenge of Detecting Synthetic Manipulations in ID Documents" (DeepID 2025)\footnote{\url{https://deepid-iccv.github.io/}}. This competition focuses on injection attacks and digital manipulation.

\section{Performance Metrics for Evaluating Computational Models}
\label{sec:metrics}

This section outlines the key performance metrics used in the reviewed literature to assess both the quality of synthetically generated identity document images and the effectiveness of Presentation Attack Detection (PAD) systems. The consistent application of standardised metrics is crucial for comparability and for advancing the field.

\subsection{Metrics for Presentation Attack Detection Evaluation}
These metrics assess the PAD system's ability to distinguish between bona fide and attack presentations (PAs). The ISO/IEC 30107-3 standard \cite{ISO-IEC-30107-3-PAD-metrics-2023} provides a foundational framework for this evaluation, and tools such as PyPAD \cite{gonzalez_forged_2025} have been developed to facilitate the computation and reporting of these metrics in a standardised manner. The selection of an appropriate decision threshold, $\tau$, which maps the system's output scores to binary decisions (bona fide or attack), is critical, as it directly influences the reported error rates. This threshold is often varied to analyse the trade-off between different types of errors.

\subsubsection{Bona Fide Presentation Classification Error Rate (BPCER)}
The BPCER \cite{ISO-IEC-30107-3-PAD-metrics-2023} measures the proportion of bona fide presentations that are incorrectly classified as presentation attacks by the system at a specific operating threshold $\tau$. It reflects the inconvenience caused to legitimate users.

Let $N_{BF}$ be the total number of bona fide presentations evaluated. Let $N_{BF \rightarrow PA}(\tau)$ be the number of these bona fide presentations that are incorrectly classified as attacks when the decision threshold is set to $\tau$ and $RES_{i}$ takes identical values to those of the $BPCER$ metric. The $BPCER$ is defined as:

\begin{equation}\label{eq:bpcer}
    BPCER=\frac{\sum_{i=1}^{N_{BF}}RES_{i}}{N_{BF}}
\end{equation}

These metrics effectively measure the degree to which the algorithm confuses presentations of attack images with bona fide images and vice versa. The APCER and BPCER metrics depend on a decision threshold.
\textbf{Interpretation:} A lower BPCER value is desirable, indicating that fewer genuine presentations are incorrectly rejected, leading to better system usability for legitimate users.

\textbf{Operational BPCER Values:} In practice, PAD systems are often evaluated at specific operational points. Metrics such as BPCER\textsubscript{10}, BPCER\textsubscript{20}, and BPCER\textsubscript{100} are frequently reported. These represent the BPCER value when the decision threshold $\tau$ is set such that the Attack Presentation Classification Error Rate (APCER) is at a predefined low level, for example, 10\%, 5\%, or even 1\% (often denoted as $BPCER$ at $APCER=X\%$). 

For instance, BPCER\textsubscript{100} might refer to the BPCER when APCER is fixed at 1\% (or a similarly stringent value; the specific APCER target for BPCER\textsubscript{100} should be explicitly stated if not clear from context or common usage in the cited literature). These metrics provide insight into system performance under specific security requirements. The threshold $\tau$ is typically determined by analysing the Detection Error Tradeoff (DET) curve of the system.

\subsubsection{Attack Presentation Classification Error Rate (APCER)}
The APCER \cite{ISO-IEC-30107-3-PAD-metrics-2023} measures the proportion of presentation attacks that are incorrectly classified as bona fide presentations by the system at a specific operating threshold $\tau$. It reflects the system's vulnerability to attacks. The standard typically reports the worst-case APCER across all considered attack types.

Equation~\ref{eq:apcer} details how to compute the APCER metric, in which the value of $N_{PAIS}$ corresponds to the number of attack presentation, where $RES_{i}$ for the $i$th image is $1$ if the algorithm classifies it as an attack presentation, or $0$ if it is classified as a bona fide presentation (real image).

\begin{equation}\label{eq:apcer}
    {APCER_{PAIS}}=1 - (\frac{1}{N_{PAIS}})\sum_{i=1}^{N_{PAIS}}RES_{i}
\end{equation}

The overall $APCER$ is then defined as the maximum $APCER$ across all evaluated attack types:
\begin{equation}
\label{eq:apcer}
\text{APCER}(\tau) = \max_{j \ge 1} \{\text{APCER}_j(\tau)\}
\end{equation}
\textbf{Interpretation:} A lower $APCER$ value is desirable, indicating that the system is better at detecting and rejecting attacks, thus enhancing security.

\subsubsection{Equal Error Rate (EER)}
The EER \cite{ISO-IEC-2382-37-220330} represents the error rate at the specific operating threshold $\tau^*$ where the system's performance in misclassifying bona fide presentations is equal to its performance in misclassifying attack presentations. In the context of PAD, this is the point where $BPCER(\tau^*)$ = $APCER(\tau^*)$. The threshold $\tau^*$ is found by varying $\tau$ across the range of possible scores.

The $EER$ value is determined when both error rates are equal:

\begin{equation}
\label{eq:eer}
EER = BPCER(\tau^*) = APCER(\tau^*)\quad 
\end{equation}

where $\tau^*$ is such that $BPCER(\tau^*)$ = $APCER(\tau^*)$.

In practice, an exact threshold where the rates are perfectly equal might not exist due to discrete score distributions; in such cases, EER is often estimated through interpolation or by finding the closest achievable point on the DET curve.

\textbf{Interpretation:} EER provides a single value summarising the overall discriminative ability of the PAD system, balancing the trade-off between falsely rejecting genuine users (usability in terms of inconvenience, related to BPCER) and falsely accepting attacks (security, related to APCER). A lower EER indicates better overall system performance, signifying stronger separation between bona fide and attack distributions.

\subsubsection{Average Ranking}

An average ranking, $AV_{Rank}$, has been proposed in the IJCB competition in 2024 and 2025 \cite{tapia_first_2024, tapia_second_2025} for ID-card PAD to determine the winning team. A weighting factor is selected for each BPCER@APCER to increase the metric's contribution in the most challenging operational points based on the threshold. Specifically, BPCER\textsubscript{10} was weighted by $0.2$, BPCER\textsubscript{20} by $0.3$, and BPCER\textsubscript{100} by $0.5$. The team with the lowest $AV_{Rank}$ won the competition. This metric weighted the BPCER\textsubscript{10,20,100} as follows to emphasise high security applications:
\vspace{-0.3cm}

\begin{equation}\label{eq:avrank}
\scriptstyle
    AV_{Rank}=BPCER_{10}\times0.2+ BPCER_{20}\times0.3 + BPCER_{100}\times0.5
\end{equation}

\subsection{Metrics for Image Synthesis Evaluation}
These metrics evaluate the quality and realism of images generated by models, such as GANs, by comparing them to real pristine images, captured from genuine ID-cards. In the context of PAD for identity documents, the quality of synthetically generated Presentation Attack Instruments (PAIs) is paramount. Effective synthetic PAIs should be realistic enough to enable PAD models to learn discriminative features of bona fide samples rather than artefacts of the generation process itself. Therefore, robust evaluation of synthetic image quality is crucial for developing reliable training datasets that can improve the generalisation capabilities and robustness of PAD systems.

\subsubsection{Fréchet Inception Distance (FID)}

FID \cite{FID} is widely used to evaluate the quality of synthetic or semi-synthetic images \cite{Joshi-SyntheticDataSurvey-PAMI-2024} produced by generative models through measuring the similarity between the distribution of generated images and the distribution of real images in a feature space. Specifically, it calculates the Fréchet distance (a measure of similarity between curves that takes into account the location and ordering of the points along the curves, adapted for probability distributions) between two multivariate Gaussians fitted to the activations of a pre-trained InceptionV3 network \cite{heusel_gans_2017}.

\textbf{Calculation Steps:}
\begin{itemize}
    \item Features (embeddings, typically 2048-dimensional from the `pool3` layer) are extracted for a set of real images ($r$) and a set of generated images ($g$) using a pre-trained InceptionV3 network.
    \item The mean vector ($\mu$) and covariance matrix ($\Sigma$) are computed for the features of the real images ($\mu_r, \Sigma_r$) and the generated images ($\mu_g, \Sigma_g$).
    \item The $FID$ score is then calculated using the means and covariances of the two distributions.
\end{itemize}

The $FID$ is formally defined as:
\begin{equation}
\label{eq:fid}
FID = \left\| \mu_r - \mu_g \right\|_2^2 + {Tr}\left( \Sigma_r + \Sigma_g - 2\left( \Sigma_r \Sigma_g \right)^{1/2} \right)
\end{equation}

where:
\begin{itemize}
    \item $\mu_r, \mu_g$ are the mean vectors of the InceptionV3 features for the real and generated image sets, respectively.
    \item $\Sigma_r, \Sigma_g$ are the covariance matrices of the InceptionV3 features for the real and generated image sets, respectively.
    \item $\left\| \cdot \right\|_2^2$ denotes the squared $L2$ norm (Euclidean distance) between the mean vectors.
    \item $\text{Tr}(\cdot)$ represents the trace of a matrix (the sum of the elements on the main diagonal).
    \item $(\Sigma_r \Sigma_g)^{1/2}$ denotes the matrix square root of the product of the covariance matrices.
\end{itemize}

\textbf{Interpretation:} A lower $FID$ score indicates that the distribution of features from generated synthetic images is similar to the distribution of features derived from real pristine images, suggesting higher quality and realism in the generated images. An $FID$ score of $0$ would indicate identical distributions.

\subsubsection{Learned Perceptual Image Patch Similarity (LPIPS)}
LPIPS \cite{LPIPS}, also known as "perceptual loss," measures the perceptual similarity between two image patches, $x$ and $x_0$. Unlike pixel-wise metrics like $MSE$ or $SSIM$, $LPIPS$ aims to better reflect human perception by comparing deep features extracted from pre-trained convolutional neural networks (e.g., AlexNet, VGG, SqueezeNet) \cite{zhang_unreasonable_2018}. The metric is calibrated using human perceptual judgments.

The LPIPS distance $d(x, x_0)$ is calculated by:
\begin{itemize}
    \item Extracting deep feature stacks from multiple layers ($L$) of a network $F$ for both image patches $x$ and $x_0$.
    \item Unit-normalizing the features in the channel dimension ($\hat{y}^l, \hat{y}_0^l$).
    \item Computing the difference between the normalised features for each layer $l$.
    \item Scaling the differences channel-wise using learned weights $w_l$. These weights are trained to make the resulting distance correlate well with human similarity judgments.
    \item Calculating the squared L2 distance for each spatial location $(h,w)$.
    \item Averaging the distances spatially across height $H_l$ and width $W_l$.
    \item Summing the results across all considered layers $l$.
\end{itemize}

The $LPIPS$ metric is formally defined as:
\begin{equation}
\label{eq:lpips}
d(x, x_0) = \sum_l \frac{1}{H_l W_l} \sum_{h,w} \left\| w_l \odot \left( \hat{y}^l_{hw} - \hat{y}^l_{0hw} \right) \right\|_2^2
\end{equation}
where:

\begin{itemize}
    \item $d(x, x_0)$ is the LPIPS distance between image $x$ and reference image $x_0$.
    \item $l$ indexes the layers used from the feature extractor network $F$.
    \item $\hat{y}^l_{hw}, \hat{y}^l_{0hw}$ are the channel-normalized feature activations at layer $l$, spatial position $(h,w)$ for images $x$ and $x_0$.
    \item $H_l, W_l$ are the height and width of the feature map at layer $l$.
    \item $w_l$ is a vector of learned weights applied channel-wise for layer $l$, optimized to match human perception.
    \item $\odot$ denotes element-wise multiplication (Hadamard product).
    \item $\left\| \cdot \right\|_2^2$ is the squared L2 norm.
\end{itemize}

\textbf{Interpretation:} A lower LPIPS score indicates higher perceptual similarity between the two images $x$ and $x_0$. A score close to 0 suggests the images are nearly indistinguishable perceptually.

\subsubsection{VGG-Loss}
The VGG-Loss \cite{VGGloss}, often referred to as Content Loss or Feature Reconstruction Loss in the context of perceptual metrics, evaluates image similarity based on high-level feature representations extracted from a pre-trained VGG network \cite{johnson_perceptual_2016}. Instead of comparing pixels directly, it compares the feature maps generated by one or more layers of the VGG network (typically VGG16 or VGG19) for a generated image $\hat{y}$ and a target or ground-truth image $y$.

The VGG feature reconstruction loss based on layer $j$ is defined as the normalised squared Euclidean distance between the feature maps:

\begin{equation}
\label{eq:vgg_loss}
l_{VGG/j}(\hat{y}, y) = \frac{1}{C_j H_j W_j} \left\| \phi_j(\hat{y}) - \phi_j(y) \right\|_2^2
\end{equation}

where:

\begin{itemize}
    \item $l_{\text{VGG},j}(\hat{y}, y)$ is the VGG loss comparing the generated image $\hat{y}$ and the target image $y$, using layer $j$.
    \item $\phi_j(\cdot)$ denotes the activation map (features) extracted from the $j$-th layer of the pre-trained VGG network for a given input image.
    \item $\hat{y}$ is the generated image, and $y$ is the target (ground-truth) image.
    \item $\left( C_j, H_j, W_j \right)$ are the number of channels, height, and width of the feature map $\phi_j$, respectively.
    \item $\left\| \cdot \right\|_2^2$ represents the squared L2 norm (sum of squared differences across all elements).
\end{itemize}

\textbf{Interpretation:} A lower VGG-Loss indicates greater similarity between the high-level feature representations of the generated image $\hat{y}$ and the target image $y$. This implies that the images share similar content and perceptual characteristics as captured by the chosen VGG layer(s). It is often used as a loss function during the training of generative models to encourage perceptual realism.

\section{Datasets}
\label{sec:datasets}
\begin{figure*}[]
    \centering
    \includegraphics[width=1\linewidth]{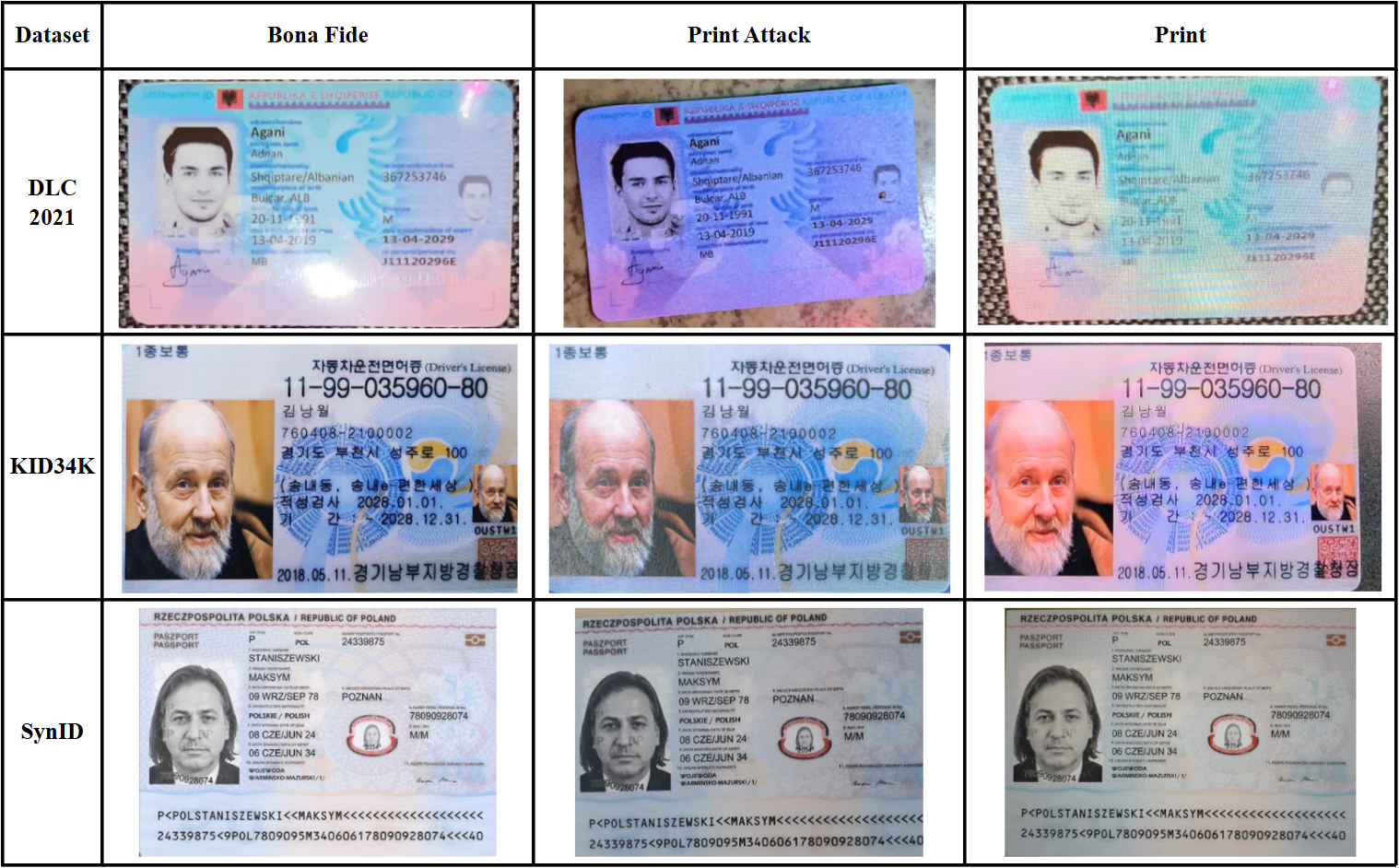}
    \caption{Illustrative examples from diverse public datasets for PAD task (DLC2021 \cite{v_document_2022}, KID34K \cite{park_kid34k_2023}, Syn-IDPASS \cite{tapia_synid_2025}).}
    \label{fig:id_examples} 
\end{figure*}

The management of controlled datasets is of utmost importance for training purposes and to obtain benchmark results based on fair and unbiased analysis. This section develops in detail datasets widely used by the scientific community for PAD tasks related to identity documents, in Figure \ref{fig:id_examples} you can see some instances that compose some of these datasets. It is essential to highlight that the examples from DLC2021 and KID34K do not represents real cases on face portraits because the face image is not ICAO-9303 compliant with respect to the requirement pose angle (no rotation) and background.

\subsection{MIDV Dataset Family}
The Mobile Identity Document Video (MIDV) dataset family represents a sustained, evolutionary effort to address the scarcity of public data for identity document analysis, a need driven by legal restrictions such as the GDPR \cite{EU2016}. Instead of using authentic documents, the MIDV series is based on the creation and capture of document mock-ups, enabling research across a range of document analysis and Presentation Attack Detection (PAD) tasks.

\textbf{MIDV-500 and MIDV-2019: Foundations and Robustness Testing.} The first dataset, \textbf{MIDV-500} \cite{arlazarov_midv-500_2019}, was conceived as a foundational resource for \textbf{general video-based identity document analysis}. Its objective was to establish a baseline for tasks such as document detection, facial recognition, and OCR of text fields. The dataset contains 500 video clips (15,000 annotated frames) of 50 document types. Capture sessions were performed with older devices (Apple iPhone 5, Samsung Galaxy S3) in Full HD resolution ($1080 \times 1920$) under controlled conditions that lacked significant lighting variations or severe projective distortions. To address these limitations, \textbf{MIDV-2019} \cite{bulatov_midv-2019_2020} was introduced, with the primary task of \textbf{testing the robustness} of algorithms under adverse conditions. This extension added 200 video clips in Ultra HD resolution ($2160 \times 3840$), captured with modern smartphones (Apple iPhone XS Max, Samsung Galaxy S10), deliberately introducing severe projective distortions and very low-light conditions.

\textbf{MIDV-2020: Benchmark for Comprehensive Analysis.} The most significant leap came with \textbf{MIDV-2020} \cite{bulatov_midv-2020_2022}, designed as a comprehensive benchmark for \textbf{evaluating complete identity document analysis systems}. Its goal was to overcome the lack of data variability by introducing $1,000$ unique physical documents (from 10 European document types), each with distinct synthetic data. This multimodal resource, the largest of its kind with $72,409$ annotated images, includes $1,000$ videos (captured with iPhone XR and Samsung S10 at 60 fps), $1,000$ photographs, and $2,000$ scans (from Canon LiDE scanners), allowing for the evaluation of end-to-end pipelines from document localisation to data extraction under diverse capture conditions.

\textbf{Specialised Extensions: Multi-script OCR and Hologram Detection.} Recognizing linguistic limitations, \textbf{MIDV-LAIT} \cite{chernyshova_midv-lait_2021} was developed for the specific task of \textbf{facilitating the training and evaluation of multi-script OCR systems}. It contains 3,600 images from 180 unique documents (from 14 countries) with complex scripts such as Perso-Arabic (Naskh and Nastaliq), Thai, and eleven Indic scripts, captured with an iPhone X Pro. Finally, \textbf{MIDV-Holo} \cite{koliaskina_midv-holo_2023} addressed a critical area of PAD, with the primary objective of \textbf{detecting physical security features in video}. This dataset, captured with an iPhone 12 and a Samsung S10, was created specifically for the task of hologram verification. It includes 300 video clips of "Utopia" documents with real custom holograms and 400 clips of presentation attacks (e.g., copies without holograms, pseudo-holograms) under various lighting conditions to highlight optically variable effects.

The MIDV dataset family has been fundamental for public research in identity document analysis. Its evolution demonstrates a continuous response to community needs, progressing from a basic dataset to collections rich in data variability, capture modalities, scripts, and physical security features. The public availability of these datasets is their greatest strength. However, their primary limitation is inherent in their design: all documents are mock-ups. They lack the complexity of substrates, inks, and covert security features of genuine documents in circulation. Although MIDV-Holo introduces holograms, these are custom designs and do not replicate the specific holograms used by governments. Furthermore, while the document designs are based on real templates, there is no guarantee of strict adherence to the requirements from the ICAO-9303 standard.

\subsection{The Document Liveness Challenge (DLC2021)}
The DLC2021 dataset \cite{v_document_2022} offers a specialised resource for the forensic analysis of video-captured identity documents, with a strong emphasis on Presentation Attack Detection (PAD). A key distinction of DLC2021 is its use of physical prop documents from the MIDV-2020 collection as its \textit{bona fide} samples. This approach allows the dataset to focus on specific forensic challenges, particularly those related to the physical properties of the document substrate.

\textbf{Dataset Composition and Diversity.} The dataset consists of 1,424 video clips of 10 different identity document types, including ID-cards and passports from Albania, Azerbaijan, Spain, Estonia, Finland, Greece, Latvia, Russia, Serbia, and Slovakia. This provides a degree of geographical diversity, although it is limited to European and nearby country templates. In compliance with GDPR, all documents are synthetic, with artificially generated personal data and photographs. The dataset models three main types of PAs, representing common, low-cost threats:
\begin{itemize}
    \item Unlaminated colour paper copies.
    \item Unlaminated grayscale paper copies.
    \item Screen recaptures from various LCD monitors.
\end{itemize}
It is important to note that the bona fide documents are the laminated versions from MIDV-2020, making lamination detection a central task and one of the dataset's main contributions.

\textbf{Capture Conditions and Realism.} To simulate real-world mobile verification scenarios, clips were captured using two popular smartphones (Apple iPhone XR and Samsung S10) at different resolutions ($1920\times1080$ and $3840\times2160$) and frame rates (30 and 60 fps). The capture process deliberately introduced significant environmental challenges, including varied lighting conditions (natural, artificial, coloured, low-light, and flash), deep shadows, and partial occlusions with fingers or other objects. The use of video is a primary strength, as it allows for the analysis of temporal artefacts, compression effects, and dynamic changes in reflection patterns that are key for PAD.

The main contribution of DLC2021 is its focus on the physical characteristics of document forgery. By providing videos of laminated originals alongside unlaminated printed copies, it directly facilitates research on using specular reflection analysis as a liveness cue, a technique that is difficult to study without such a specific dataset. The dataset is publicly available on Zenodo under a CC BY-SA 2.5 license, representing a significant benefit for reproducible research. However, certain limitations should be considered. The \textit{bona fide} documents, while high-quality prints, are still mock-ups. They lack the complex, covert, and optically variable security features (e.g., advanced holograms, OVI inks) present in genuine circulating documents. Moreover, while the document templates are based on real examples, their strict adherence to the ICAO-9303 standard is not guaranteed. Consequently, models trained on DLC2021 may excel at detecting simpler print and replay attacks but may not generalise to detecting forgeries that mimic sophisticated physical security features.

\subsection{Korean Identity Document 34K (KID34K)}
The KID34K dataset \cite{park_kid34k_2023} was developed to address a very specific and prevalent use case: \textbf{fraud detection in online identity document verification for the South Korean market}. Its primary goal is to train AI models capable of differentiating between images captured from a genuine, physical ID-card and those that are recaptures from a screen or a printed copy, which are the most common types of presentation attacks.

\textbf{Dataset Composition and Diversity.} As its name suggests, the dataset consists of a total of \textbf{34,662 images}. It focuses exclusively on two of the most used identity documents in South Korea: the resident registration card and the driver's license. For the dataset's creation, 46 synthetic identities were generated, resulting in 82 unique physical ID-cards (37 registration cards and 45 driver's licenses). The dataset is labelled into three main categories, defining the capture scenario:
\begin{itemize}
    \item \textbf{Bona fide:} A photographic capture of the physical plastic card.
    \item \textbf{Screen:} A recapture of the ID document image displayed on an electronic device's screen (smartphone, tablet, etc.).
    \item \textbf{Print:} A recapture of a version of the ID document printed on paper.
\end{itemize}

\textbf{Sample Generation and Acquisition.} A fundamental aspect of KID34K is that its \textit{bona fide} samples are digital templates but captures of \textbf{physical plastic cards} manufactured specifically for the dataset. To ensure privacy, all personal information is synthetic. AI-generated facial images ("This person does not exist") were used, and textual data (names, dates) were created in such a way that a Korean observer would recognise them as inauthentic, although the layout and format of the fields are not visually similar to real documents. The image acquisition was performed under a wide variety of conditions to simulate real-world scenarios, using different models of smartphones and tablets, varying lighting conditions, and changing the brightness and size of the document image in screen and print attacks.

The main strength of KID34K lies in its \textbf{practical approach and capture realism} for a specific market. Using physical plastic cards as the basis for the \textit{bona fide} class provides a higher level of realism than datasets using only digital templates, allowing models to learn subtle material characteristics. Furthermore, being public and using fully synthetic data, it resolves the privacy issues that have prevented the publication of other similar datasets. However, its limitations are noteworthy. Its \textbf{geographical and document-type scope is extremely limited}, raising questions about the generalizability of models trained on it to documents from other countries or with different designs. The attack vectors are restricted to replay (print and screen) and \textbf{do not include digital composite attacks} (e.g., photo or text alteration) or more sophisticated physical forgeries. As the cards are synthetic, they lack the physical and optical security features (OVDs, holograms, etc.) that are crucial for detecting high-quality fakes.

\subsection{Synthetic ID and Travel Documents (SIDTD)}
The SIDTD dataset \cite{boned_synthetic_2024} is introduced as a direct extension of MIDV-2020, with the primary goal of facilitating the \textbf{training and evaluation of Presentation Attack Detection (PAD) systems} focused on common, low-complexity document forgeries. Unlike other datasets, SIDTD establishes a clear distinction: it treats the MIDV-2020 samples as the \textit{bona fide} class and generates a new set of digitally altered samples to represent the attack class (\textit{forged}).

\textbf{Composition and Attack Types.} SIDTD reuses the 1,000 documents of 10 European nationalities from MIDV-2020 as its authentic dataset. The main contribution of the dataset is the generation of 1,222 new forged document templates using two \textbf{composite} attack techniques:
\begin{itemize}
    \item \textbf{Crop \& Replace:} This involves swapping specific regions (like the signature or photo) between two documents of the same type, introducing a slight offset to generate detectable edge artefacts.
    \item \textbf{Inpainting:} This implies digitally removing information from a text field and rewriting it with a different font, simulating an alteration of the original content.
\end{itemize}
These methods simulate simple digital forgeries that do not require expert knowledge, representing a common type of threat in online verification systems.

\textbf{Sample Generation and Acquisition.} To increase the realism of the attacks, 191 of the digitally generated forged templates were printed on photo paper with a laser printer (HP Colour LaserJet E65050), laminated, and manually cut. Subsequently, videos of these physically forged documents were recorded using a wide range of smartphones (over 25 different models), ensuring great diversity in image quality, lighting conditions, and backgrounds. Finally, 7,214 clips (frames) were extracted from these videos, which, together with the 68,409 clips from the \textit{bona fide} documents of MIDV-2020, form the video component of the dataset.

SIDTD is a valuable resource for the community, as it is one of the few public datasets explicitly designed for the task of PAD in identity documents. Its main strength is providing a large volume of data (images and videos) to train models capable of detecting common digital manipulations. The publication of the code to generate attacks encourages reproducibility. However, it shares the limitations of its base, MIDV-2020. The \textit{bona fide} documents are mock-ups and lack the advanced physical security features of real documents. By extension, the attacks are limited to digital alterations on these mock-ups, without addressing the simulation of physical security features (holograms, OVI inks) or more sophisticated attacks. Adherence to the ICAO-9303 standard is also not guaranteed.

\subsection{IDNet}
The IDNet dataset \cite{guan_idnet_2024} emerges as a response to the scale and diversity limitations of existing public datasets. Its primary objective is to provide a \textbf{large-scale benchmark of synthetic identity documents for training and evaluating fraud detection and analysis tools}, with a strong focus on privacy preservation. To date, IDNet is the largest publicly available dataset of synthetic identity documents.

\textbf{Dataset Composition and Diversity.} The scale of IDNet is its main differentiator, with a total of \textbf{837,060 synthetic document images}. The dataset covers 20 document types from 10 U.S. states and 10 European countries. For each document type, 5,979 unique \textit{bona fide} samples are generated. One of its most significant contributions is the diversity of the model's presentation attacks, generating six fraudulent variants for each \textit{bona fide} sample:
\begin{itemize}
    \item \textbf{Crop \& Move} and \textbf{Inpaint \& Rewrite}, replicating the attacks from SIDTD.
    \item \textbf{Face Morphing}, merging two distinct faces into a single portrait image.
    \item \textbf{Portrait Substitution}, replacing the original photo with one that does not meet quality requirements.
    \item \textbf{Complex Text Replacement}, altering not only the content but also the style, font size, and background colour of personal identifier information (PII) fields.
    \item \textbf{Mixed Attacks}, combining several of the previous fraud techniques.
\end{itemize}

\textbf{Sample Generation and Acquisition.} IDNet introduces a novel few-shot, AI-assisted synthetic data generation pipeline. The process begins with the automated creation of templates from a few real document samples, using a diffusion model (Stable Diffusion 2.0) to remove personal information. Subsequently, an LLM (ChatGPT-3.5-turbo) is used to generate textual metadata, ensuring semantic constraints are met. To compose the final document, a Bayesian optimiser automatically adjusts hyperparameters to maximise the structural similarity (SSIM) with the original sample.

The main strength of IDNet is its \textbf{massive scale and unprecedented diversity of modelled fraud patterns}. By including complex attacks like face morphing that directly intersect with PII fields, the dataset is positioned as an invaluable resource for research in privacy-preserving PAD. Its few-shot generation pipeline is highly efficient. However, as a purely synthetic dataset, the generated \textit{bona fide} samples lack the complex physical and optical security features of authentic documents. The modelled attacks, though diverse, are digital in nature. Therefore, while IDNet is powerful for detecting a wide range of digital manipulations, its applicability for detecting high-end physical forgeries is limited.

\subsection{Syn-IDPASS}
The Syn-IDPASS dataset \cite{tapia_synid_2025} is presented as a solution to address one of the main weaknesses of existing synthetic datasets: the lack of realism and non-compliance with international standards. Its primary goal is to provide a dataset of \textbf{high-fidelity synthetic passports, generated in compliance with ICAO-9303 requirements}, to improve the training and evaluation of PAD systems.

\textbf{Dataset Composition and Diversity.} Syn-IDPASS focuses exclusively on passports and, in its initial version, features three nationalities: Spain, Poland, and Portugal. The core of the dataset consists of 3,000 \textit{bona fide} images (1,000 per country) generated via a hybrid method that combines empty passport templates with synthetically generated textual and biometric data. A differentiating aspect is that the entire generation process, from facial images to the Machine Readable Zone (MRZ), follows the ICAO-9303 specifications. The dataset models two types of physical presentation attacks:
\begin{itemize}
    \item \textbf{Print Attacks:} The generated templates are printed on 600 dpi glossy photo paper to simulate the material of a real passport.
    \item \textbf{Screen Attacks:} The generated images are displayed on a laptop screen and recaptured.
\end{itemize}
In total, the dataset comprises 9,000 images: 3,000 \textit{bona fide}, 3,000 print attacks, and 3,000 screen attacks.

\textbf{Sample Generation and Acquisition.} The generation of \textit{bona fide} samples is a meticulous process starting from photoshop templates. Subject data is generated from culturally appropriate dictionaries, and facial images are selected and rigorously filtered to meet ICAO-9303 criteria. For the attacks, the generated templates are printed or displayed on-screen and manually recaptured, ensuring a variety of capture conditions.

The main strength of Syn-IDPASS is its \textbf{emphasis on compliance with the ICAO-9303 standard}, producing synthetic documents with a higher level of structural and biometric realism than many other public datasets. This makes it a potentially more effective resource for training PAD models that need to learn the subtle features of compliant documents. However, the dataset has clear limitations: its geographical scope is very narrow, and it does not include composite (digital) attacks. Most critically, like other synthetic datasets, the \textit{bona fide} documents lack physical security features (OVDs, etc.), which are crucial for detecting sophisticated forgeries.

\subsection{RSCID}
The Recaptured Student Card Image Dataset (RSCID) \cite{chen_domain-agnostic_2022} was developed to address a critical and specific challenge: \textbf{domain generalisation in document recapture detection}. Its primary objective is to provide a benchmark for evaluating the robustness of PAD algorithms when faced with significant variations in capture devices, printing technologies, and physical substrates.

\textbf{Dataset Composition and Diversity.} The dataset consists of a total of \textbf{1,104 images}, divided into 132 genuine captured images and 972 recaptured attack samples \cite{chen_domain-agnostic_2022}. 
It focuses exclusively on student ID-cards from five different university templates, which were synthetically designed to avoid privacy issues. The core of RSCID is structured into two main subsets, $D_1$ and $D_2$, each collected using a distinct set of hardware to deliberately create different data domains. The dataset models several types of recapture attacks:
\begin{itemize}
    \item \textbf{Print-and-Scan Recaptures:} The primary attack, where a captured image of a genuine document is printed and then re-acquired.
    \item \textbf{Display-and-Capture Recaptures:} Attacks where the document image is displayed on an LCD screen and recaptured with a camera.
    \item \textbf{Substrate Variation Attacks:} Recaptures using high-quality glossy photo paper to test robustness against different printing materials.
\end{itemize}

\textbf{Sample Generation and Acquisition.} A distinguishing feature of RSCID is the nature of its \textit{bona fide} source documents. Instead of simple paper mock-ups, the documents were first designed using Adobe CorelDRAW and then professionally \textbf{manufactured on acrylic plastic} to better simulate the physical properties of real cards \cite{chen_domain-agnostic_2022}. 

To generate the dataset, these physical plastic cards were first captured using various devices to create the genuine samples. Subsequently, these captured images were printed on standard $120 g/m^2$ office paper and re-acquired using a wide array of scanners and smartphone cameras to create the recaptured attack samples. 

The dataset's key contribution to domain generalisation is its use of two distinct hardware sets: Dataset $D_1$ was created with a combination of consumer-grade smartphones, scanners, and an inkjet printer, while $D_2$ employed a different set of higher-quality devices, including 48-megapixel camera phones and high-resolution printers, to test generalisation across different quality levels \cite{chen_domain-agnostic_2022}.

The main strength of RSCID lies in its \textbf{explicit focus on the domain shift problem}, providing a publicly available and well-documented resource for benchmarking model generalisation across different hardware. The use of physical plastic cards as the foundation for the process adds a layer of material realism compared to datasets based on paper mock-ups. However, the dataset has clear limitations. Its scope is restricted to student ID-cards and does not represent the complexity of official government-issued documents. As the documents are synthetic, they \textbf{lack advanced physical and optical security features} (e.g., holograms, OVI). Furthermore, the dataset does not claim compliance with standards like ICAO-9303, and the attack vectors are centered on recapture, \textbf{without including digital composite attacks} such as photo splicing or text inpainting.

\subsection{FantasyID}
The FantasyID dataset \cite{FantasyID} is introduced as a direct solution to the limitations of existing public datasets, with the primary objective of facilitating research in \textbf{detecting digital manipulations of ID-documents}. Its key differentiator is that, unlike previous works, it provides truly pristine \textit{bona fide} samples, free from prior digital alterations (such as the removal of 'specimen' watermarks), and \textbf{is publicly available for commercial use}.\\

\textbf{Dataset Composition and Diversity.} FantasyID is built upon \textbf{13 unique design templates} that emulate identity documents in 10 different languages, including Arabic, Chinese, French, Persian, and Russian, providing significant cultural and linguistic diversity. The dataset features 362 unique cards designed with randomised but realistic personal data and faces of real people sourced from public databases, thereby avoiding the biases of AI-generated faces. The modelled attacks are exclusively \textbf{digital manipulations} that simulate an injection attack scenario, where an attacker modifies a captured image before submitting it to a verification system. These include:
\begin{itemize}
    \item \textbf{Face Swapping:} Manipulations performed with state-of-the-art tools such as InSwapper and Facedancer to replace the original facial image.
    \item \textbf{Text Inpainting:} Alterations to text fields using diffusion models like Textdiffuser2 and DiffSTE to change sensitive information such as names or dates.
\end{itemize}

\textbf{Sample Generation and Acquisition.} The creation process for the \textit{bona fide} samples is a core strength of the dataset. The 362 unique digital cards, designed from scratch using Creative Commons licensed materials, were physically printed using a \textbf{specialised card printer (Evolis Primacy 2)} onto plastic cards to emulate the materiality of real documents. 
Subsequently, these physical cards were captured using three different devices (an iPhone 15 Pro, a Huawei Mate 30, and an office scanner) to generate a total of \textbf{1,086 \textit{bona fide} images} that simulate a realistic \textit{Know Your Customer} (KYC) scenario. The attack samples are, therefore, digital manipulations performed on these already captured \textit{bona fide} images.\\

The main strength of FantasyID is that it provides the research community with a public dataset featuring truly pristine simulated \textit{bona fide} samples and a permissive commercial use license, removing the biases present in datasets derived from altered specimens. Its linguistic diversity and the use of real faces on physically printed plastic cards make it a highly realistic resource for training and evaluating digital manipulation detectors. However, the dataset has explicit limitations. 

The documents \textbf{do not strictly adhere to the ICAO-9303 standard}. Furthermore, its focus is exclusively on \textbf{digital manipulation (injection) attacks}, excluding simpler physical presentation attacks such as screen or low-quality print recaptures, which are also common threat vectors.

\subsection{FakeIDet}
The FakeIDet-db dataset \cite{FakeIDet} is introduced as a direct response to a major bottleneck in the field: the absence of public data from real identity documents. Its primary objective is to propose and validate a \textbf{privacy-preserving patch-based methodology for fake ID detection}, thereby enabling the publication of data derived from authentic documents. Its most significant contribution is being a \textbf{publicly available database containing patches extracted from real Spanish ID documents} and their corresponding physical attacks.

\textbf{Dataset Composition and Diversity.} The public dataset, named \textbf{FakeIDet-db}, contains a total of 48,400 image patches. These are derived from an original set of \textbf{30 real Spanish} national identity cards, belonging to 30 distinct subjects and covering three different design templates to increase variability. The dataset models two high-quality physical Presentation Attack Instrument (PAI) species:

\begin{itemize}
    \item \textbf{Print Attacks:} The real documents were scanned at 600 DPI, printed on regular paper, and subsequently laminated to enhance their realism.
    \item \textbf{Screen Attacks:} The same scanned images were displayed on a high-pixel-density screen (254 PPI) and carefully recaptured to avoid artefacts such as moiré patterns.
\end{itemize}

\textbf{Sample Generation and Acquisition.} The FakeIDet methodology is founded on a \textbf{privacy-preserving patch extraction} process. The 30 real documents and the 60 generated physical attacks were photographed with a low-end smartphone to simulate a realistic capture scenario. Prior to patch extraction, the full images undergo an anonymisation process where sensitive information (text, face, signature) is obscured. The public database only includes patches from the \textbf{pseudo-anonymised and fully-anonymised} versions. From these images, patches of various sizes ($128\times128$, $64\times64$, and $32\times32$) are extracted, which, on their own, do not contain identifiable sensitive information.

The main strength of FakeIDet-db is that it is the \textbf{first public resource derived from real identity documents}, offering an unprecedented level of realism for research and establishing an innovative solution to the dilemma between privacy and data utility. The modelling of high-quality physical attacks, such as laminated prints, also presents a challenge that is closer to real-world scenarios. However, the dataset has significant limitations. Its scope is \textbf{geographically very limited} (only Spanish IDs) and it is based on a small number of subjects (30). Furthermore, it focuses solely on physical presentation attacks, \textbf{excluding digital composite manipulations} (such as splicing or inpainting). Finally, by distributing only patches, it precludes research on methods that require analysis of the global context or the full layout of the document. 

A summary of the datasets is depicted in Table \ref{tab:datasets_synthesis}.

\begin{table*}[]
\centering
\caption{Analysis and Comparison of Public Datasets for Identity Document PAD}
\label{tab:datasets_synthesis}
\resizebox{\textwidth}{!}{%
\begin{tabular}{ccccc}
\hline
\textbf{Dataset} &
  \textbf{Primary Goal} &
  \textbf{Document Scope} &
  \textbf{Attack Types Modeled} &
  \textbf{Bona Fide Nature} \\ \hline
\begin{tabular}[c]{@{}c@{}}\textbf{MIDV Family}\\  \cite{arlazarov_midv-500_2019, bulatov_midv-2019_2020, bulatov_midv-2020_2022, koliaskina_midv-holo_2023, chernyshova_midv-lait_2021}\end{tabular} &
  \begin{tabular}[c]{@{}c@{}}Evolving benchmark for general\\ analysis, robustness, multi-script\\ OCR, and hologram detection.\end{tabular} &
  \begin{tabular}[c]{@{}c@{}}50+ types (IDs, passports, etc.)\\ covering European, Perso-Arabic,\\ Thai, and Indic scripts.\end{tabular} &
  \begin{tabular}[c]{@{}c@{}}\textit{Base:} None\\ \textit{MIDV-Holo:} Print copies,\\ pseudo-holograms (400 clips).\end{tabular} &
  \begin{tabular}[c]{@{}c@{}}Laminated paper mock-ups.\\ (1.9k docs, 76k images/clips).\\ MIDV-Holo includes custom\\ physical holograms (300 clips).\end{tabular} \\ \hline
\begin{tabular}[c]{@{}c@{}}\textbf{DLC2021}\\ \cite{v_document_2022}\end{tabular} &
  \begin{tabular}[c]{@{}c@{}}Forensic analysis with a focus on\\ physical substrate properties, \\ (e.g., lamination detection).\end{tabular} &
  \begin{tabular}[c]{@{}c@{}}10 European types (IDs and passports)\\ from MIDV-2020.\end{tabular} &
  \begin{tabular}[c]{@{}c@{}}Print (unlaminated colour/grayscale),\\ screen recaptures (1,134 clips).\end{tabular} &
  \begin{tabular}[c]{@{}c@{}}Laminated mock-ups from\\ MIDV-2020 (290 clips).\end{tabular} \\ \hline
\begin{tabular}[c]{@{}c@{}}\textbf{KID34K} \\  \cite{park_kid34k_2023}\end{tabular} &
  \begin{tabular}[c]{@{}c@{}}PAD for the South Korean market,\\ detecting print and screen attacks.\end{tabular} &
  2 Korean types (resident card, driver's license). &
  \begin{tabular}[c]{@{}c@{}}Print (7.2k), screen recaptures\\ (13.7k images).\end{tabular} &
  \begin{tabular}[c]{@{}c@{}}Custom-manufactured physical\\ plastic cards (13.7k images).\end{tabular} \\ \hline
\begin{tabular}[c]{@{}c@{}}\textbf{SIDTD}\\  \cite{boned_synthetic_2024}\end{tabular} &
  \begin{tabular}[c]{@{}c@{}}PAD for common, low-complexity\\ digital composite attacks.\end{tabular} &
  10 European types (from MIDV-2020). &
  \begin{tabular}[c]{@{}c@{}}Digital Composite (Crop \& Replace,\\ Inpainting) (1.2k templates, 7.2k clips).\end{tabular} &
  \begin{tabular}[c]{@{}c@{}}Digital templates from\\ MIDV-2020 (1k templates, 68k clips).\end{tabular} \\ \hline
\begin{tabular}[c]{@{}c@{}}\textbf{IDNet}\\ \cite{guan_idnet_2024}\end{tabular} &
  \begin{tabular}[c]{@{}c@{}}Large-scale benchmark for privacy,\\ preserving fraud detection.\end{tabular} &
  \begin{tabular}[c]{@{}c@{}}20 types (10 U.S. states, \\ 10 European countries).\end{tabular} &
  \begin{tabular}[c]{@{}c@{}}Digital Composite, Face Morphing,\\ Portrait Substitution, Text Replacement,\\ Mixed Attacks (717k images)\end{tabular} &
  \begin{tabular}[c]{@{}c@{}}AI-generated digital templates\\ via few-shot pipeline (119.5k images).\end{tabular} \\ \hline
\begin{tabular}[c]{@{}c@{}}\textbf{Syn-IDPASS}\\ \cite{tapia_synid_2025}\end{tabular} &
  \begin{tabular}[c]{@{}c@{}}High-fidelity synthetic dataset, \\ generation compliant with ICAO-9303\\ standards.\end{tabular} &
  3 European passport types. &
  Print (3k), screen recaptures (3k). &
  \begin{tabular}[c]{@{}c@{}}Digital templates generated\\ 3 countries (9k images).\end{tabular} \\ \hline
\begin{tabular}[c]{@{}c@{}}\textbf{RSCID}\\ \cite{chen_domain-agnostic_2022}\end{tabular} &
  The Recaptured Student Card Image Dataset &
  BID recaptured &
  Print, Screen, Plastic &
  \begin{tabular}[c]{@{}c@{}}132 genuine captured images, \\ No ICAO 9303.\end{tabular}
    \\ \hline
\begin{tabular}[c]{@{}c@{}}\textbf{FantasyID}\\ \cite{FantasyID}\end{tabular} &
  Synthetic Dataset for Injection Attack &
  Multiple ID-cards, and a resident permit &
  Print, Capture and Digital manipulations &
  \begin{tabular}[c]{@{}c@{}}Digital templates generated \\ fake templates (1086 images).\end{tabular}
    \\ \hline
\begin{tabular}[c]{@{}c@{}}\textbf{FakeIDet}\\ \cite{FakeIDet}\end{tabular} &
  Bona fide patches for privacy protection &
  ID-cards from Spain &
  Print, Screen &
  48,400 patches from 30 bona fide images \\ \hline
\end{tabular}%
}
\end{table*}

\section{Discussion}
\label{sec:discussion}

The systematic analysis presented in this review illuminates a central paradox within the current landscape of PAD for identity documents. While the field employs increasingly sophisticated Deep Learning methodologies, its progress is fundamentally constrained by the foundational challenge of data scarcity. Unlike established PAD research for biometric modalities such as face \cite{abdullakutty_review_2021} or fingerprint recognition \cite{ametefe_advancements_2025}, the domain of identity documents remains a comparatively nascent area. This review confirms that its primary research thrusts are not merely about designing deeper networks, but about architecting smarter, more data-efficient solutions to navigate a landscape defined by limited data access.

Our findings reveal an architectural evolution precipitated by this data constraint, transitioning from general-purpose classifiers to specialized, forensically-focused models. While the literature is dominated by established CNN architectures such as MobileNet \cite{sandler_mobilenetv2_2019, gonzalez_forged_2025} and ResNet variants \cite{chen_domain-agnostic_2022}, more innovative approaches acknowledge that the primary challenge lies not in a deficit of feature extraction capability, but rather in the difficulty of learning generalizable, fraud-related features from heterogeneous and scarce data. This has precipitated a distinct trend towards task decomposition, exemplified by multi-stage systems that employ dedicated models for digital and physical attacks \cite{gonzalez_hybrid_2021}.

Moreover, there is a growing focus on micro-level artefact analysis, wherein methods operate on pixels or patches to detect subtle forensic traces introduced by printing or recapture processes \cite{mudgalgundurao_pixel-wise_2022, chen_domain-agnostic_2022}, and even extracting information observable only in the frequency domain in search of generalizable forensic traces \cite{li_two-branch_2023, chen_multi-modal_2024, chen_spectral_2024}. This forensic granularity represents a key theme, demonstrating a strategic shift away from holistic image classification towards a more targeted, evidence-based detection paradigm.

The recent exploration of Vision Transformers and Foundation Models \cite{tapia_can_2025, FakeIDet} represents the latest step in this trajectory, but their full potential for capturing both local forensic details and global document context remains an open and promising area of investigation. The FM appears as a new insight for achieving generalisation capabilities with the same number of images available. Open a new insight to create a PAD that supports multiple countries. 

A significant consequence of the data scarcity issue is a clear divide between academic research and industrial applications. Our review found that a majority of high-performing models are trained and validated on extensive, private datasets, a practice that, while effective, severely hinders reproducibility and objective benchmarking. This limitation creates a bifurcation in the field: academic progress is primarily benchmarked against public datasets composed of mock-ups or synthetic data (e.g., MIDV \cite{bulatov_midv-2020_2022}, IDNet \cite{guan_idnet_2024}, DLC-2021 \cite{v_document_2022}), whereas industrial systems evolve on proprietary, real-world data. 

The notable performance gap between these two areas, especially evident in academic competitions such as the one at IJCB \cite{tapia_first_2024}, indicates that current public benchmarks may not fully capture the complexities of real-world scenarios. This disconnect poses a risk, as it limits the academic community's ability to assess and improve the robustness and fairness of commercially deployed systems.

The turn towards synthetic data generation \cite{benalcazar_synthetic_2023, markham_open-set_2024} can be interpreted as a significant strategic response to this data access crisis. Techniques based on Generative Adversarial Networks (GANs) have proven valuable for augmenting training sets and enabling research in a privacy-preserving manner. However, our analysis suggests that while synthetic data helps, it does not solve the challenge. A persistent "reality gap" remains, where models risk overfitting to artefacts of the generation process rather than learning bona fide attack cues \cite{benalcazar_synthetic_2023}. The observed disconnect between perceptual quality metrics like FID and a model's downstream task performance \cite{markham_open-set_2024} underscores this challenge: visually plausible fakes are not always forensically effective training samples.

This very limitation of synthetic data has catalysed the recent and powerful shift towards a new paradigm: the adoption of large-scale Foundation Models (FMs) \cite{tapia_can_2025, FakeIDet}. This trend represents a fundamental change in strategy: if creating perfectly realistic data for the model is intractable, the alternative is to adapt a model that has already learned rich visual representations from web-scale data. The success of models like DinoV2, which demonstrate remarkable generalisation capabilities even when fine-tuned on small, privacy-preserving patches from real documents \cite{FakeIDet}, signals that leveraging vast, pre-existing knowledge is a more effective path to robustness than attempting to synthesise it from scratch. This approach not only addresses the generalisation problem but also enables novel privacy-preserving frameworks, marking a pivotal moment in the field's evolution.

In synthesis, the field of PAD for identity documents is characterised by a sophisticated and adaptive response to its core data constraint. The research trajectory has evolved from applying general deep learning architectures to developing specialised forensic systems to attempting to create new realities via synthetic generation and has now arrived at leveraging the vast knowledge embedded in foundation models. This journey highlights a dynamic and resourceful research community. The path forward, therefore, lies not just in algorithmic refinement, but in pioneering innovative data strategies — be it through privacy-preserving techniques, federated learning frameworks \cite{Fed}, or more advanced synthetic generation — that can finally bridge the gap between academic research and the demands of globally deployed, trustworthy verification systems.

\section{Conclusions} \label{sec:conclusions} This systematic review has mapped the AI 'arms race' in identity document security, conclusively demonstrating that the field is defined by a central paradox: while Deep Learning (DL) and Foundation Models (FMs) offer unprecedented defensive sophistication, its academic progress is fundamentally shackled by the scarcity of realistic training data.

Our analysis has traced the methodological evolution this scarcity has forced: from generic CNN architectures, through specialized forensic models at the micro-artefact level \cite{chen_multi-modal_2024, mudgalgundurao_pixel-wise_2022}, to the effort to create new data via synthetic generation \cite{benalcazar_synthetic_2023, markham_open-set_2024}. However, we identify three critical gaps that define the current boundaries of research and must be the focus for the next generation of studies:

The Reality Gap: An abyssal performance divide exists between models trained on extensive, private industry datasets and those trained on existing public benchmarks (composed of mock-ups or synthetic data). The results from competitions such as IJCB PAD-ID-card \cite{tapia_first_2024, tapia_second_2025} empirically confirm this: our current academic proxies fail to capture real-world domain complexity, thereby impeding reproducibility and the translational relevance of research.

The Synthetic Utility Gap: Synthetic data generation (e.g., GANs) has been established as a direct response to data scarcity. However, as evidenced in \cite{markham_open-set_2024}, a perilous disconnect exists between perceptual fidelity (e.g., FID metrics) and forensic utility. Models risk overfitting to the artefacts of the generation process rather than learning the artefacts of the attack process, creating a false sense of security.

The Paradigm Gap: The limitations of models used in synthetic data generation to close the reality gap have catalyzed the most significant paradigm shift identified in this review: the adoption of Foundation Models (FMs). Pioneering studies \cite{tapia_can_2025, FakeIDet} demonstrate that it is more effective to adapt the vast pre-trained knowledge of models like DinoV2 to scarce data (and even to privacy-preserving patches \cite{FakeIDet}) than to attempt to synthesise a complete forensic reality from scratch.

In synthesis, the field of PAD for identity documents must pivot. The focus must migrate from the mere optimisation of accuracy on synthetic datasets to the radical pursuit of real-world generalisation, data efficiency, and privacy-preserving evaluation frameworks.

\section{Future Work: A Prescriptive Research Agenda} \label{sec:future_work} Based on the critical gaps identified, this review proposes a prescriptive research agenda designed to guide the scientific community toward robust, generalisable, and deployable solutions.

\subsection{From Mock-ups to Federated and Privacy-Preserving Benchmarks} The most urgent need is not simply 'more data,' but 'better data.' Instead of relying on mock-ups (e.g., MIDV \cite{bulatov_midv-2020_2022}) or purely synthetic templates (e.g., IDNet \cite{guan_idnet_2024}), research must focus on frameworks that permit training and evaluation on real data without compromising privacy. Future work must radically expand the patch-based approach of FakeIDet \cite{FakeIDet} to use real data without the ethical risks of identity reconstruction, and explore Federated Learning (FL) methodologies. This would allow multiple institutions (e.g., banks, governments) to train a robust PAD model on their private data without any sensitive data ever leaving their servers.

\subsection{Exploiting Foundation Models for Forensic Generalization} FMs \cite{tapia_can_2025} have demonstrated superior potential for generalization. Future research must move beyond using them as mere feature extractors (backbones). The objective must be to investigate whether FMs can be fine-tuned to explicitly disentangle forensic traces \cite{chen_multi-modal_2024} from the document's semantic features. Can a model like DinoV2 learn a universal representation of a 'print artefact' or a 'recapture moiré pattern' that is invariant to the document type, country, or language?

\subsection{Forensically-Aware Explainable AI (XAI)} The 'black box' nature of DL models is a critical liability in the high-security domain. It is not sufficient for a model to classify a document as 'fake'; we must know why. Future research must develop XAI methods specifically designed for document forensic analysis. Instead of generic heatmaps (e.g., Grad-CAM), we need techniques that highlight specific pixel-level artefacts (e.g., halftone pattern inconsistencies, print noise \cite{li_two-branch_2023}, or text manipulation traces \cite{Zhao}) that justify the model's decision to a human analyst.

\subsection{Dedicate metrics} At present, there are no established metrics for measuring the similarity of synthetic images that include both facial portraits and demographic information text. As outlined in this study, most existing metrics are adaptations of those used for synthetic face recognition. This area presents opportunities for further development and improvement.

In the era of generative AI, trust in digital identity can no longer be taken for granted; The level of trust must be verified algorithmically and robustly. Without a fundamental shift towards real-world generalisation and proactive threat assessment, as outlined in this review, the global identity infrastructure remains critically vulnerable.

{\small
\bibliographystyle{ieee}
\bibliography{Arxiv-version/references}
}

\end{document}